\documentclass[manuscript]{acmart}
\pdfoutput=1
\usepackage{siunitx}
\usepackage{booktabs} 
\usepackage{listings} 
\usepackage{booktabs} 
\usepackage[ruled]{algorithm2e} 
\usepackage{indentfirst} 
\usepackage{paralist}
\usepackage{graphicx} 
\usepackage[]{siunitx} 
\usepackage{comment}
\usepackage{mathtools}
\usepackage{amsmath}
\usepackage{graphicx,wrapfig,lipsum} 
\usepackage[font=small,labelfont=bf,tableposition=top]{caption} 
\usepackage{graphicx}
\usepackage{etex}
\usepackage{capt-of}
\usepackage{xfrac}
\usepackage{color} 
\usepackage{gensymb} 
\usepackage{subfig}
\usepackage{textcomp}
\usepackage[ruled]{algorithm2e} 

\usepackage{color}
\definecolor{darkblue}{rgb}{0.0, 0.0, 0.55} 

\AtBeginDocument{%
  \providecommand\BibTeX{{%
    \normalfont B\kern-0.5em{\scshape i\kern-0.25em b}\kern-0.8em\TeX}}}

\copyrightyear{2021}
\acmYear{2021}
\setcopyright{acmlicensed}\acmConference[UIST '21]{The 34th Annual ACM Symposium on User Interface Software and Technology}{October 10--14, 2021}{Virtual Event, USA}
\acmBooktitle{The 34th Annual ACM Symposium on User Interface Software and Technology (UIST '21), October 10--14, 2021, Virtual Event, USA}
\acmPrice{15.00}
\acmDOI{10.1145/3472749.3474823}
\acmISBN{978-1-4503-8635-7/21/10}

\begin{document}

%

\def\startupVoltage{100}
\title{MARS: Nano-Power Battery-free Wireless Interfaces for Touch, Swipe and Speech Input} 
 
%


\author{Nivedita Arora}
\email{nivedita.arora@gatech.edu}
\affiliation{%
  \institution{School of Interactive Computing, Georgia Institute of Technology}
  \country{USA}
}

\author{Ali Mirzazadeh}
\affiliation{%
   \institution{School of Interactive Computing, Georgia Institute of Technology}
  \country{USA}}
\email{alimirz@gatech.edu}

\author{Injoo Moon}
\affiliation{%
\institution{School of Interactive Computing, Georgia Institute of Technology}
  \country{USA}
}
\email{imoon8@gatech.edu}

\author{Charles Ramey}
\affiliation{%
\institution{School of Interactive Computing, Georgia Institute of Technology}
  \country{USA}
}
\email{cramey7@gatech.edu} 

\author{Yuhui Zhao}
\affiliation{%
\institution{School of Interactive Computing, Georgia Institute of Technology}
  \country{USA}
 }
\email{yzhao343@gatech.edu}

\author{Daniela C. Rodriguez}
\affiliation{%
\institution{School of Interactive Computing, Georgia Institute of Technology}
  \country{USA}
}
\email{dxrodriguez@gatech.edu}

\author{Gregory D. Abowd}
\affiliation{%
\institution{Dept. of Electrical and Computer Engineering, Northeastern University}
  \country{USA}
 }
\affiliation{%
\institution{School of Interactive Computing, Georgia Institute of Technology}
  \country{USA}
}
\email{g.abowd@northeastern.edu}

\author{Thad E. Starner}
\affiliation{%
  \institution{School of Interactive Computing, Georgia Institute of Technology}
  \country{USA}}
\email{thad@gatech.edu}

\renewcommand{\shortauthors}{Arora, et al.}

\begin{abstract} 
Augmenting everyday surfaces with interaction sensing capability that is maintenance-free, low-cost ($\sim$ \$1), and in an appropriate form factor is a challenge with current technologies. MARS (\textbf{M}ulti-channel \textbf{A}mbiently-powered \textbf{R}ealtime \textbf{S}ensing) enables battery-free sensing and wireless communication of touch, swipe, and speech interactions by combining a nanowatt programmable oscillator with frequency-shifted analog backscatter communication. A zero-threshold voltage field-effect transistor (FET) is used to create an oscillator with a low startup voltage ($\sim$500 mV) and current ($<2uA$), whose frequency can be affected through changes in inductance or capacitance from the user interactions. Multiple MARS systems can operate in the same environment by tuning each oscillator circuit to a different frequency range. The nanowatt power budget allows the system to be powered directly through ambient energy sources like photodiodes or thermoelectric generators. We differentiate MARS from previous systems based on power requirements, cost, and part count and explore different interaction and activity sensing scenarios suitable for indoor environments.
\end{abstract}


\ccsdesc[500]{Human-centered computing~Interaction devices}
\ccsdesc[500]{Hardware~Communication hardware, interfaces and storage}
\ccsdesc[500]{Hardware~Power and energy}

\keywords{Interaction, Low-voltage, Low power, Flexible Electronics, Backscatter, Tangible, Wireless, Sensing}

\maketitle

\section{Introduction}\label{Intro}
Post-it\textregistered\ sticky notes are used to convey reminders on surfaces in the environment. Imagine similar \textbf{computationally-enabled stickers that can be easily attached to such surfaces to sense human interactions and wirelessly convey this information to computational assistants}. We present MARS, (\textbf{M}ulti-channel \textbf{A}mbiently-powered \textbf{R}ealtime \textbf{S}ensing), a prototyping platform that enables creation of speech, touch and swipe gesture-based wireless sticker interfaces. MARS stickers can be used as an extended microphone for a smart home equipped with a smart voice assistant (\autoref{fig:application_scenario}A). It can be embedded in a printed conference brochure to communicate questions asked by the attendee to the auditorium's amplification system so that the attendee can be heard (\autoref{fig:application_scenario}B). Like post-it notes, these MARS interaction stickers can be made into many different shapes and sizes to attach to other objects. For example, a rectangular interaction sticker attached under each item on a printed food menu (\autoref{fig:application_scenario}C) can allow a customer to order the item by simply sliding their finger across it. This remote control capability can be used in other scenarios such as lighting controls (\autoref{fig:application_scenario}D). In addition, the MARS interaction sticker can be designed to convey its meaning visually; for example, it can mimic the design of a game controller (\autoref{fig:application_scenario}E)
 or reveal multiple-choice options on a desk in a classroom for interactive polling.

\begin{center}
\vspace{-0.1in}
\begin{figure}[!th]
\includegraphics[width=\columnwidth]{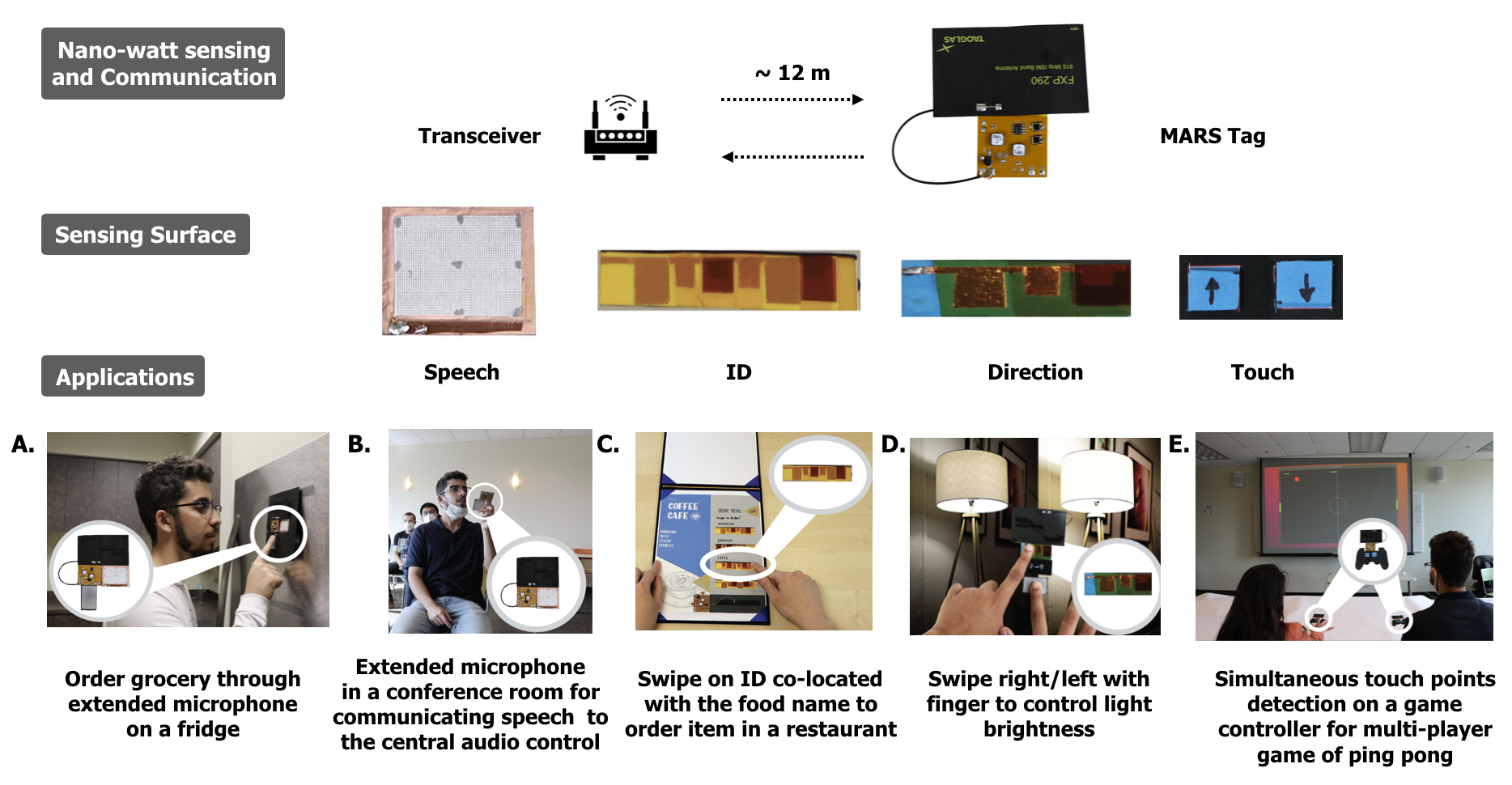} 
\caption{\textbf{MARS Interfaces:} 
\textbf{A.} 
Battery-free wireless microphone, activated by touch, extends the hearing range of smart home devices.
\textbf{B.}
A similar microphone embedded in an attendee's conference brochure transmits the attendee's voice to the auditorium's speaker system so that they may ask a question.
\textbf{C.} 
Capacitive swipe ID sensors associated with each item on an interactive menu wirelessly transmit a customer's order to the restaurant's kitchen. \textbf{D.} 
Capacitive slider dims a remote lamp. 
\textbf{E.} 
Wireless touch interface stickers can designed in shapes that indicate their use, such as game controllers (shown) or classroom polling systems.} 
\label{fig:application_scenario}
\end{figure}
\vspace{-0.23in}
\end{center}

Practical wide-scale creation and deployment of MARS sticker interfaces require addressing several design constraints. Like an everyday sticky note, the MARS stickers should be self-sustainable in their operation and should not require regular \textbf{power maintenance}. One way to do that is to remove batteries from the system and instead use ambient power for operation. Since we are targeting MARS interaction stickers for indoor home or office use, the power harvesting needs should be satisfied in illumination  $\sim$ 500 lux \cite{OSHALight}, temperatures in the range of 68-76° F and humidity controlled to be in the range of 20\%-60\% \cite{OSHATemp}. Next, the stickers should be \textbf{low-cost} (e.g., under \$1) such that they can be used and discarded (even recycled) when not needed. Removing batteries, often the most expensive component in low power electronics, would significantly lower component cost, as would  \textit{minimizing the harvester size}. \textit{Reducing circuit part count and complexity} would further lower the assembly costs. Finally, the MARS interaction sticker \textbf{form factor} should be flat and small enough to be easily applied and unobtrusive, while large enough to be seen, understood, and accommodate sensing interactions. Our goal is for MARS tags to be comparable to the 3-inch by 3-inch Post-it\textregistered\ note. To achieve this, all sub-parts of the tags (sensing, communication, and power harvesting) need to conform to the form factor constraints. 
The required components for sensing and communication should be thin or show the future possibility of being printable.

\vspace{-0.1in}
\subsection*{Contributions and Limitations}
We achieved the power, cost, form factor design parameters for MARS sticky notes by creating a circuit for \textit{frequency modulation (FM) -based backscatter communication} \cite{ranganathan2018rf,wang2017fm}. 
Leveraging the Zero-$V_{th}$ MOSFET \cite{junior2012zero}, we build a modified version of the traditional LC Clapp oscillator to have low-startup voltage (500mV), current ($\sim$ 2uA), and power consumption ($<1uW$) up to a frequency of 1 Mhz (\autoref{sec: tag_design_comm}). The resonant frequency of the Modified Clapp Oscillator (MCO) is controlled by the inductance (L) and capacitance values (C), which is exploited for inductance, capacitance and voltage based analog sensing. This technique allows MARS to enable a diverse set of interfaces (speech, slider, ID, and touch) that can augment objects/surfaces in the environment by simply placing a sticker. We use our previous work on a self-powered paper microphone, SATURN \cite{arora2018saturn}, for speech sensing and design novel capacitance-based direction and identity sensors. With minimal startup power, MARS tags can communicate sound up to 9m and other interactions up to 12m (in monostatic configuration). We demonstrate that MARS interfaces can be powered by two photodiodes harvesting ambient light in an office environment or a thermoelectric generator converting power from heat flow from the user's fingertip (\autoref{power_harvesting}).  MARS interactive stickers are built using 9 components (2 active and 7 passive) and less expensive power harvesters than most previous systems (\autoref{power_harvesting}, \autoref{sec:case_study}). 

While MARS interactive sticker circuit design could lead to IoT interfaces at significantly less cost than current systems (\autoref{sec:formfactor}, \autoref{sec:case_study}), they also hold some limitations. Replication of MARS tag frequency is affected by the parasitic reactances and the quality factor of the passive components (\autoref{replicate}). While the MARS sticker tags are built using minimal components, there is more work to be done to make them printable (\autoref{sec:formfactor}). Several more strategies with respect to range (\autoref{sec:range_discussion}), placement of tags (\autoref{sec:tag_placement}), power harvesting alternatives (\autoref{sec:altPH}) can be adopted to increase the robustness of a real-world deployment. Finally, further work needs to be done to address the user privacy concerns (\autoref{sec:privacy}).

\section{Background and Related Work} 
Previous work falls into three key categories. We first explore everyday object- and surface-based interaction sensing technologies with respect to the system design parameters mentioned earlier (\autoref{interaction_sensing}). Second, we review system design and technology trends in the domain of battery-free wireless interaction sensing (\autoref{batteryfree_interaction_sensing}). Third, we describe  existing low-voltage and low-power transistor analog circuits.  We explore their applicability to battery-free wireless interaction sensing systems like MARS (\autoref{lv_tech}). Finally, we provide a case study comparing MARS to a commercial product (Amazon Dash) and the closest project in the literature with similar capabilities (RF-bandaid).

\subsection{Object- and surface-based interaction sensing}\label{interaction_sensing}
In Mark Wieser's vision of ubiquitous computing,  computing could be brought into the physical world through tabs, pads, and boards \cite{weiser1999computer}.  As this technology improved, \textit{touch and speech} became topics of great interest in the HCI community. Acoustic sensors (microphone \cite{harrison2008scratch, laput2015acoustruments}, capacitive \cite{olwal2018braid, savage2012midas}, light \cite{wimmer2010flyeye, butler2008sidesight}, piezoelectric \cite{ono2013touch}, and resistive \cite{holman2014sensing}) have been used to recognize touch as well as fine-grained gestures. Yet they typically depend on complicated circuitry in the form of a micro-controller, ADC, and power management that limit their suitability for scaling to a large number of low-cost interactive objects \cite{islam2019zygarde}. In addition, the batteries typically required to support active sensors can render the form factor unsuitable to being incorporated into or onto everyday objects. Conductive paint \cite{zhang2018pulp,zhang2018wall++} and flexible circuits \cite{wang2019flextouch} have been proposed to enable touch interactions at scale. While these proposed sensing parts are flexible, complete systems often have shortcomings in power, complexity of circuit, and form factor.

Speech interactions with smart home assistants  \cite{yamazaki2012home, davidoff2006principles} have been a focus of commercial and academic research. However, such assistants are often limited to microphones in a single location where the smart hub is located. To enable users to being able to mount a remote microphone where ever needed, we previously built a flexible microphone called SATURN that can be used to sense human speech and environmental sounds \cite{arora2018saturn, arora2020saturn}. However, SATURN still required a means to support multiple wireless sound tags in the same environment. 

\vspace{-0.05in}
\subsection{Battery-free wireless interaction sensing }\label{batteryfree_interaction_sensing}

To create a battery-free wireless sensing system, the power being harvested from the environment or human interaction needs to match the system power budget. Often satisfying this requirement results in trade-off of functionality, power, form factor and circuit complexity. One of the most power consuming component of a wireless sensing system is the active generation of radio waves to communicate the sensed information; often 10-100 mW is required depending on the frequency and distance desired. To realize a self-sustainable battery-free system with active radio the power source needs to be intense (e.g., constant motion/vibration of an equipment or human, large temperature difference to generate thermal energy, close proximity to RF source \cite{paradiso2005energy}) and often bulky in form factor with complicated power management (e.g., sheet of solar panel, electromagnetic coils). Sometimes a balance can be achieved between power and functionality with clever hardware design. For example, Enocean's ECO200 battery-free wireless switches \cite{enocean} send an ID powered by the mechanical movement of pushing a button.  Sozu \cite{zhang2019sozu} can detect presence and ID by harvesting power in everyday environments. These results are encouraging yet are still limited for sensing interactions like audio and gestures that are power-intensive due to requiring a higher sampling rate and bandwidth. 

Another common battery-free communication technique is \textit{backscatter} \cite{xu2018practical}. In this scenario, an active transmitter transmits a carrier radio wave in the vicinity of the tag, while the tag backscatters the carrier wave with modulation (e.g., amplitude, frequency, and phase) based on sensed information.
Both analog and digital techniques can be utilized to modulate the incident radio waves in backscatter communication devices. While digital backscatter communication (e.g., RFID \cite{yeager2008wisp}) is generally considered more robust, analog communication requires an order of magnitude less power, and simpler circuitry than digital circuits \cite{ranganathan2018rf,zhang2016enabling, arora2018zeusss}. While amplitude modulation backscatter is noisy and does not support simultaneous operation of tags \cite{brooker2013lev}, frequency shifted/modulated-backscatter communication allows for wireless communication from different types of analog sensor tags on different channels. Here, we exploit the interaction of the human body with the tag's tank circuit to re-tune the FM resonator for simultaneously sensing and transmitting the interaction, simplifying the circuit and lowering the required power. Next we discuss how chipless RFID \cite{preradovic2010chipless}, UHF RFID \cite{li2015idsense, paperid}, and analogue backscatter (amplitude \cite{talla2017battery,arora2018zeusss} and frequency \cite{zhang2016enabling,ranganathan2018rf} modulation) technologies support the battery-free wireless sensing of diffrent interactions.

\paragraph{\textbf{Touch:}} ID sense \cite{li2015idsense}, PaperID \cite{paperid}, and RapID \cite{spielberg2016rapid} have demonstrated extending RFID-based single touch interaction capabilities to objects by detecting changes in backscattered signals. These methods often require separate tags for each touchpoint sensor. LiveTag \cite{gao2018livetag} showed multiple touchpoint detection by chipless RFID antenna, but was limited by the number of tags (every two meters) that must reside within the five-meter operable distance of the transmitter. Tip-Tap is an RFID tag-based wearable input technique for 2-dimensional discrete touch events, but requires wearing gloves for interaction\cite{katsuragawa2019tip}. MARS allows simultaneous detection of multiple discrete touch-points on separate channels in $<1uW$ of power with a range of 15 m. 

\vspace{-0.1in}
\paragraph{\textbf{Identity:}}
Many of the RFID-based wireless touch sensors mentioned previously can be considered wireless tags that communicate identity \cite{li2015idsense, paperid}. For example, human-powered RFID buttons harvest the power from hard-push button interactions to activate their radio to communicate a unique ID to a central server \cite{paradiso2001compact}. These events are discrete whereas, with MARS, more continuous states can be transmitted. For example, the dashes and dots of Morse code keying can be conveyed, or the rhythm of a song. Papergenerators \cite{karagozler2013paper} communicates ID through an IR LED powered by the prolonged and continuous rubbing of a triboelectric generator. With reduced power needs, MARS can communicate a touch by the user simply resting their finger on the thermoelectric part of the tag.

\vspace{-0.12in}
\paragraph{\textbf{Swipe-based gesture sensing:}}
Individual RFID tags connected in a circular or linear pattern can detect swipe gestures \cite{paperid}. RIO \cite{pradhan2017rio} detects gestures by recognizing phase changes caused by finger contact. 3D printed objects which have rotatory parts, e.g., pill bottle, prosthetic hand, anemometer, can manipulate impedance of a 3d printed antenna with their movement for direction and speed sensing \cite{iyer2018wireless, iyer20173d}. Self-sustainable active communication powered by high-energy actions (e.g., a door closing) have been proposed for direction sensing, but such techniques do not support  interactions such as speech or simple touches as they are too low in power \cite{zhang2019sozu}. Recently, Ubiquitouch  has shown FM backscatter-based touch and swipe sensing but requires a power budget of about 35 $\mu {W}$ and is powered by a solar cell of 50$cm^2$ \cite{ubiquitouch}. Such systems cannot be easily converted to a post-it like form factor or have the simplicity of MARS (<10 components), whose simplicity may lead to it being printable (\autoref{sec:formfactor}). 

\vspace{-0.12in}
\paragraph{\textbf{Sound:}} 
Battery-less wireless sensing of sound is specifically tricky since it requires a high sampling rate and communication bandwidth, both of which come at the price of power and increase in the complexity of the circuit. Often the power is high (mW), such that only batteries can satisfy the power requirement, thus effecting both form factor and cost. A hybrid analog-digital modulation scheme has been proposed to sense and transmit acoustic data, which leverages the Wireless Identification and Sensing Platform (WISP) \cite{talla2013hybrid}. Later RF-BandAid \cite{ranganathan2018rf} opened the possibility for more robust purely analog frequency-modulated backscatter communication of audio range signals for physiological monitoring in 35-150 $\mu{W}$ up to  1Mhz. Pre-encoded audio has been backscattered in ~10 $\mu{W}$ through FM modulation leveraging a simulated capacitor and a cross-coupled NMOS  relaxation oscillator circuit \cite{wang2017fm}.  Inspired by analog based FM-backscatter, we built MARS to enable sensing different types of human interactions in indoor settings, in $<1uW$ power, <10 components,  and a part cost of ~\$1 (10,000 units) (\autoref{sec:case_study}).  

\vspace{-0.05in}
\subsection{Low Voltage (LV) and Low-Power (LP) circuit designs} \label{lv_tech}
Circuit operation at reduced supply voltages is a common practice adopted to reduce the power consumption. At low supply voltage, the main constraints faced are the MOSFET device noise level and the threshold voltage ($V_{th}$) \cite{yan2000low}. Reduction in $V_{th}$ is dependent on the device technology. Higher $V_{th}$ gives better noise immunity and the lower $V_{th}$ reduces the noise margin to result in poor SNR. Several advancements in transistor device fabrication and circuit design have been accomplished to allow low $V_{th}$ without effecting the device performance -- MOSFETs operating in the sub-threshold region, bulk driven transistors , self-cascode structures, floating gate approach  and the level shifter techniques \cite{rakus2017design, yan2000low}. For MARS stickers we explore one such low-voltage threshold device, Zero-$V_{th}$ MOSFET by Advanced Linear Devices \cite{ALD110800}, for its applicability in building a low-voltage, low-power oscillator circuit (\autoref{mosfet_primer}). Lowering $V_{th}$ allows lowering of the circuit's startup voltage that in-turn enables impedance match with common low-voltage DC harvesters (e.g., photo-diode in indoor room lighting,  human-touch based thermal electric generator) without complex power management. Thus, low voltage circuits support both our aims of low-power and circuit simplicity.

\vspace{-0.1in}
\subsection{Comparative Case Study: Amazon Dash button, RF-band-aid, and MARS} \label{sec:case_study}
To better ground this paper's contributions toward the goal of interface stickers, we first calculate the amount of power available in an office environment. We then compare what energy harvesters are needed to power MARS compared to current IoT devices and recent devices reported in the literature.







\vspace{0.3in}
\begin{tabular}[b]{ccc c}\hline
 & \textbf{MARS} & \textbf{RF-Bandaid}\cite{ranganathan2018rf} & \textbf{Amazon Dash} \cr
\hline 
\textbf{Max Power} & $<$1uW & 160 uW & 300mW \cr
\hline 
\textbf{Startup Voltage} & $~$0.5V & 2.6V & 3V \cr
\hline 
\textbf{Parts Cost} & $\sim$\$1.4 & \$9.5; \$20 w/ solar; \$24 w/  & $>$\$10; \$310 w/ harvester; \\&& thin-film battery & $>$\$11.50 w/ battery \cr
\hline 

\textbf{Min Weight} & 3.5 g & 11g (solar), 6g (battery)
& 150g\cr
\hline 
\textbf{Min Surface Area} & 9$cm^2$  & 9$cm^2$  & 29$cm^2$\cr
\textbf{w/o Harvester} &(w/ dipole antenna) & (w/dipole antenna)\cr
\hline 
\textbf{Min Volume} & $\sim$0.5$cm^3$ & $\sim$0.5$cm^3$ & 73.6$cm^3$ (PCB) \cr
\hline 

\textbf{Battery} & n/a & \$15, 15 hr life (thin film) & \$1.50 in bulk; 7.6g; 3.8$cm^3$ \\&&    ST Microelectronics     & 1.8Whr AAA  \\ && EFL1K0AF39 & Energizer lithium\cr
\hline 

\textbf{Solar Harvester } &  0.2$cm^2$ & 45$cm^2$   & 10,000 $cm^2$ \\\textbf{Surface Area (200-500lux)}&2 photodiodes& 8.5 indoor solar cells & \cr
\hline 
\textbf{Total Solar Harvester Cost} & \$0.4 \cite{MARSPhotodiode}  & \$11.60  \cite{am-1417ca}  &  \$300
\cr
\hline 
\textbf{No. of Components} & 9 (2 actives, 7 passive) & 100+ active, 100+ passive
 & millions\cr
\hline 

\end{tabular}
\captionof{table}[text for list of tables]{Comparison between different wireless communication systems that could support sound, swipe, identity, and touch sensing}
\label{table:comparison}

While  illumination should be above 500 lux in an office, an interface sticker could be placed so it is not directly facing the illumination source.  To take this factor into account, we assume that the interface sticker will have at least 200 lux available, which is the value used in data tables for small solar cells used for powering calculators and other small indoor devices \cite{am-1417ca}. The total amount of power available in an office equipped with fluorescent lights and illuminated at 200-500 lux is then between $0.33 mW/cm^2$ and $0.83 W/cm^2$ (LED $0.22$ - $0.55 mW/cm^2$; incandescent $1.3$-$3.3 mW/cm^2$). Energy conversion for typical small indoor amorphous silicon solar cells is currently around 9\% \cite{NextGenSolar}. However, previous efforts such as the RF Band-Aid also require energy storage, power management circuitry and, in some cases, boost converters to achieve the required start-up voltage.  As reported, the effective power harvested on the inside of a window was $8 \mu W/cm^2$ and a much lower $0.7 \mu W/cm^2$ on a desk \cite{grosse2016exploring} (suggesting a harvester surface area need of at least $50cm^2$). 

Body heat from finger or hand is another source of energy readily available for interaction applications. A finger tip of 3$cm^2$ at  97.8 °F has 15 mW of power available through heat flow (50W/$m^2$ for body \cite{starner2004human}). Thermoelectric generators have  0.2-0.8\% efficiency \cite{lay2009thermoelectric, stevens1999optimized} for heat conversion from the body  at room temperature ($\sim$20°C) and thus can generate 30 $uW$ from finger touch. This level of power is reasonable for MARS,  especially with low voltages,  but not for RF-bandaid or Amazon Dash. 

We compare MARS to the commercial Amazon Dash product which has the capability of touch (one button but has the internal circuitry to sense multiple capacitive buttons), sound (MEMS microphone), and active radio transmission. 
The RF BandAid \cite{ranganathan2018rf} provides a state-of-the-art comparison to a system in the literature most similar to MARS. The RF BandAid  includes a microphone and senses changes in a force sensing resistor and chest strap. It also maps sensor output to frequency modulation for wireless backscatter transmission.

\autoref{table:comparison} demonstrates the benefit of MARS's extremely low energy operation in comparison to other devices with respect to parts cost and the size of solar harvester needed to power the system. While Dash’s battery should last 59 years in sleep mode (3.45uW), it lasts only 6 hours if continuously transmitting a touch or sound (300mW). Unlike MARS, every second of event transmission reduces battery life in sleep mode by 1 day. Alternatively, a common solar harvester sufficient to register a button push or audio in indoor lighting (200-500 lux) is impractical at a surface area of  10,000$cm^2$ and a cost of \$300 \cite{DashSolarCell}. Specifically, in order to match MARS’s ability for registering continuous button interactions or sound, this large size is needed to produce the required max current of 300mW. If the input was more occasional, a supercapacitor could be used to store harvested energy until it was needed, and a less expensive harvester could be used.  However, even examining commercial products with solar harvesters that do not require continuous monitoring, such as the Logitech solar keyboard (35 $cm^2$) and Phillips remote control (30 $cm^2$), shows that these devices require harvesters much larger than MARS ($0.2 cm^2$)\cite{apostolou2016comparison}.

Like MARS, the RF-BandAid can continuously monitor input, but it requires 160uW of power with a 2.6V startup voltage. 45$cm^2$ of solar harvester is needed to meet these specifications. MARS's main advantage over RF-BandAid is its 100x lower power, which allows a 200X smaller surface area harvester and 10X reduction in cost. However, another advantage is the low startup voltage, which allows a better impedance match with low voltage power harvesters without the need for power management.  This advantage reduces the number of components in the system (9 components versus the RF BandAid's $\sim$ 200 or the Amazon Dash's millions), thus lowering cost, and opens the possibility of making the entire system printable in the future, like wallpaper or house wrap  \cite{vallgaarda2007computational,abowd2020IoM}.

Like Dash or the RF-BandAid, MARS requires infrastructure, but ``ubiquity'' could be achieved by embedding MARS receivers in the power outlets or light fixtures of a building and exploiting power line communication back to a central location, such as a smart speaker. By reducing part cost, requiring a small surface area, and having a flat form factor, MARS enables a vision of interfaces placed like stickers on surfaces in the environment; a vision other methods have difficulty achieving.


\section{MARS SYSTEM DESIGN AND THEORY OF OPERATION}


Our MARS system consists of multiple MARS tags augmented onto different objects and surfaces in the environment within the transmitter and receiver communication range. The transmitter is responsible for producing a high frequency carrier radio signal (e.g., $F_{high}$ in \autoref{fig:3_system_operation})  which is incident on the tags. Leveraging a custom ultra-low power oscillator, each of the MARS tags are tuned to a unique frequency (e.g., F1-3 in \autoref{fig:3_system_operation}) much lower than the transmitter's carrier frequency. The unique resonant frequency allows each tag to communicate information in their own individual channel. For example, Tag 1 sends information at F1 shifted from $F_{high}$. Based on the interactions the tag oscillator shifts a single frequency or modulates a band of frequencies (speech). \autoref{fig:3_system_operation} shows an example of frequency shifting for interaction with Tag 1 where the sensor changes value as a sine wave. This frequency modulated sensed information signal from the oscillator is communicated in real time to the receiver by amplitude modulating it with the incoming high frequency carrier signal. This technique, where the active radio source is in the transmitter and the tag is only responsible to modulate and reflect the incoming signal, is called backscatter. Depending on the bandwidth of data (i.e., single frequency  or 8kHz audio) being communicated during the interaction, this communication protocol is called frequency shifted or modulated analog backscatter \cite{ranganathan2018rf,zhang2016enabling}. We employ this protocol to communicate speech, touch, ID, swipe information from MARS tags. 

\begin{center}
\begin{figure}[!ht]
\includegraphics[width=\columnwidth]{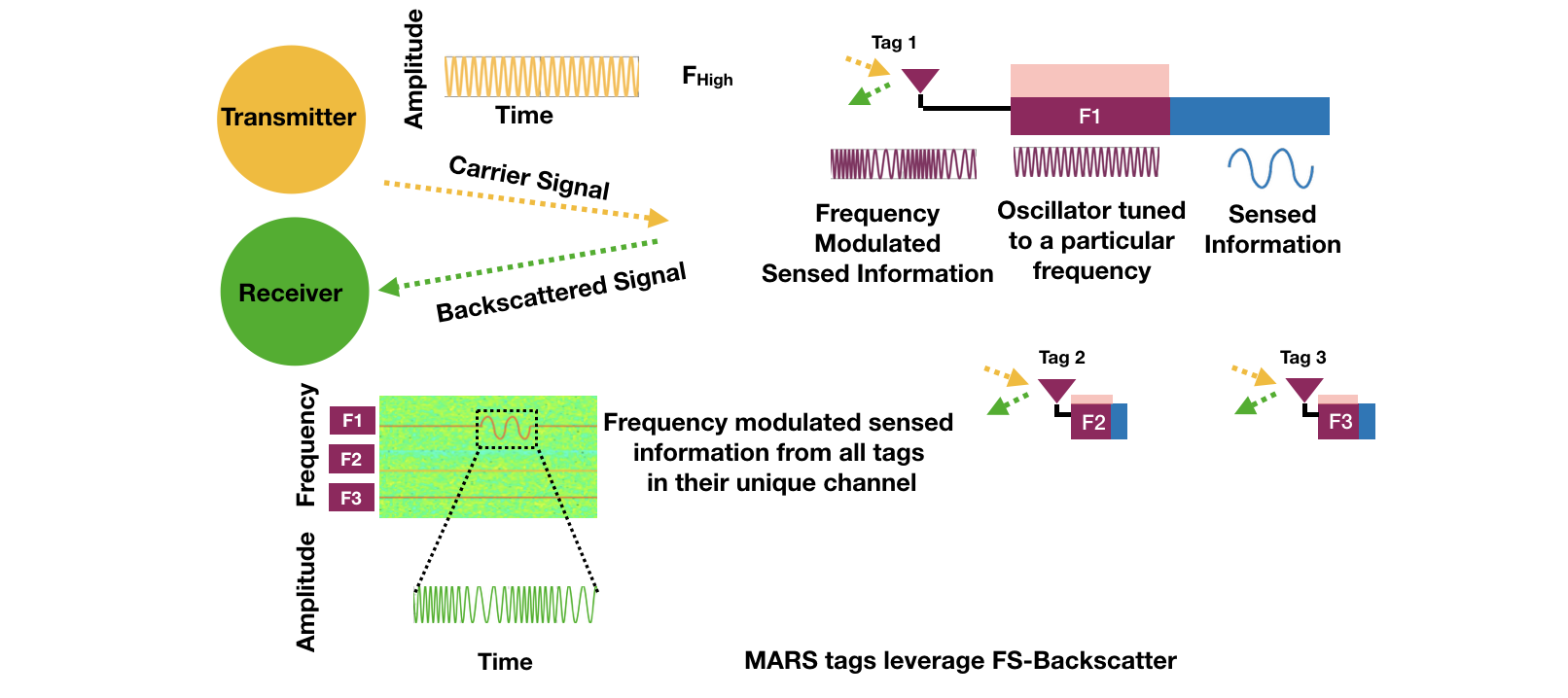}
\caption{ \textbf{MARS System based on Frequency Shifted-Backscatter:}  MARS system consisting of transmitter, receiver and multiple MARS tags in the environment. The input interactions result in change in electrical property (blue) resulting in frequency modulating the oscillator signal (magenta). This signal is backscattered  to the receiver (green) using an analog impedance switch and antenna. The receiver, receives frequency modulated information at different bands from each tag, which is demodulated back to the original sensed signal. } 
\label{fig:3_system_operation}
\end{figure}
\vspace{-0.3in}
\end{center}

One of our main contributions is the hardware design implementation of the nano-watt power budget MARS wireless sensing tag. The MARS tag (\autoref{fig:3_tag}A) consists of three main blocks as shown in  \autoref{fig:3_tag}B; (1) the \textit{communication block }(magenta), which is responsible for the FS-backscatter; (2) the \textit{power harvesting block} (red), which leverages ambient power sources (e.g., body heat, photodiode) to power the communication; and (3) the \textit{sensing block} (blue), which enables sensing of different electrical phenomenon (inductance L, capacitance C, or voltage V) to support interaction sensing.

In the following subsections we detail each of the tag blocks and their operation. We first start with a primer on Zero $V_{th}$ MOSFET, an piece of known technology that is essential for realisation of the ultra-low power and startup-voltage oscillator used in the communication sub-block. Next, we explain different communication sub-blocks -- oscillator, analog switch, and antenna. We specifically focus on a detailed description of the hardware design process and characterisation of the oscillator. This section is followed by elucidation of the role and selection of components for the analog switch and antenna. Next is a discussion on how the tag design enables sensing based on the inductance, capacitance and voltage change. Finally, we will discuss how the tag gets powered from ambient power sources like light and body heat to enable wireless sensing.  

\begin{center}
\begin{figure}[!ht]
\includegraphics[width=0.9\columnwidth]{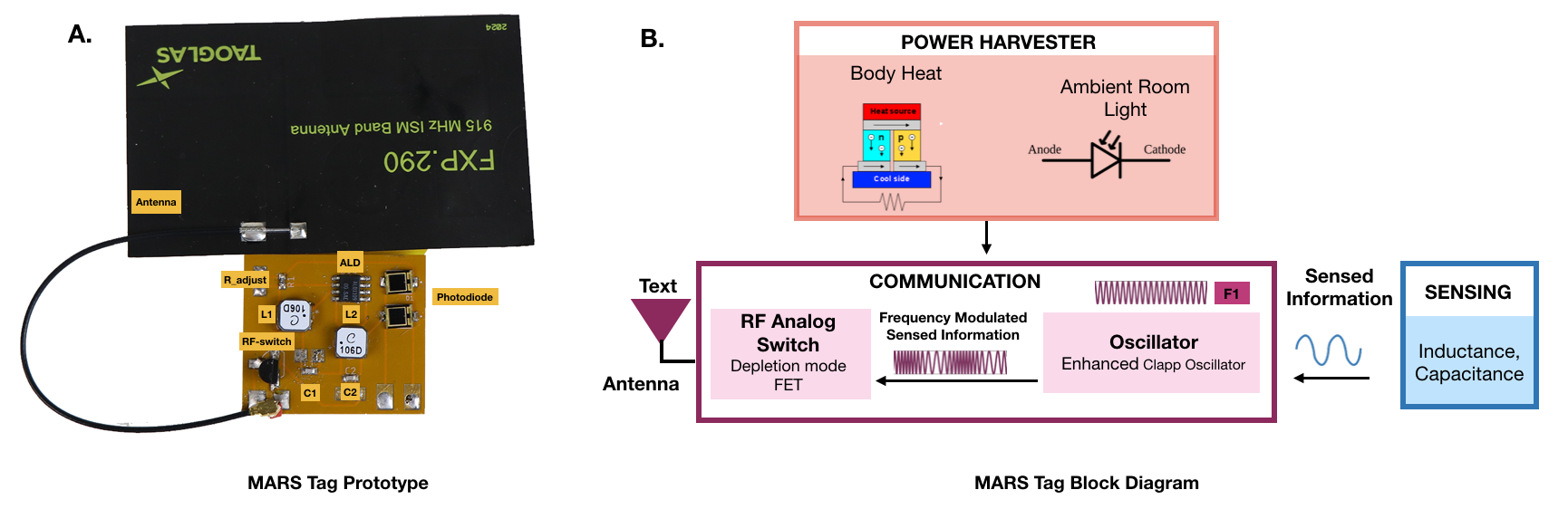}
\caption{ \textbf{MARS Tags} \textbf{A.} prototype \textbf{B.} The three major components of the MARS tag: communications, power harvester, sensing.} 
\label{fig:3_tag}
\end{figure}
\vspace{-0.3in}
\end{center}

\subsection{Primer on Zero $V_{th}$ MOSFET}
\label{mosfet_primer}

The MOSFET is a four terminal device (body, gate, source and drain) which forms a basis for modern electronics. Each MOSFET device has a threshold voltage, $V_{th}$, set during the manufacturing process, that is the minimum amount of voltage applied to the gate needed to create a conducting path between the source and drain terminals. A typical MOSFET requires a supply voltage, $V_{dd}$, which is a few 100's mV higher than $V_{th}$ to operate. This difference between $V_{dd}$ and $V_{th}$ is called the gate overdrive. Since, the overall device power requirement is quadratically proportional to the $V_{dd}$, one way to lower the circuit power consumption is to lower the $V_{dd}$. As the feature size in modern MOSFET devices scales down, the required $V_{dd}$ decreases (Dennard scaling \cite{1050511}), but the $V_{TH}$ does not scale down at an equal rate  \cite{gonzalez1997supply}. If, however, the $V_{dd}$ is lowered too much, it diminishes gate overdrive, which results in performance degradation of MOSFET and increase in the static power dissipation. A potential workaround to maintain consistent behavior with a lower supply voltage is to reduce the threshold voltage $V_{th}$ so as to maintain an adequate gate overdrive.

Inspired by applications for low-voltage (LV) and low-power (LP), analog designers have found alternative ways to maneuver around this limited voltage headroom by manufacturing specially designed MOSFETs . Zero-$V_{th}$ MOSFETs were introduced as a component in the mid 90's \cite{burr1994200}, but it was soon realized that these transistors had an exponential increase in sub-threshold current \cite{macii2004ultra} (green in  \autoref{fig:theory_of_operation_zero_vth}A) in comparison to the standard MOSFET (red in figure \autoref{fig:theory_of_operation_zero_vth}A), resulting in significantly high leakage power. Thus, Zero-$V_{th}$ MOSFETs were never received with much acceptance in the hardware community.

Recently, however, by leveraging a floating gate MOSFET design \cite{seo2004low} and propriety EPAD technology \cite{EPAD} together, manufacturers like Advanced Linear Devices have been able to reduce $V_{th}$ to $\sim$ 0 mV (e.g., ALD110800A \cite{ALD110800}) while still maintaining leakage current $\sim 1uA$. This balance results in an extremely low supply voltage, even near 100 mV, to be practical for circuits. These novel near Zero-$V_{th}$ transistors hold a great potential for nano-powered analog and RF-systems. The green region marked in (\autoref{fig:theory_of_operation_zero_vth}B) shows a overdrive voltage range of +/- 0.15V and current $<3 uA$. In our paper we exploit this region to build an ultra-low startup power and voltage oscillator as discussed in the \autoref{sec:oscillator}. Similar low power oscillators have been described previously in the literature using floating-gate MOSFET \cite{matsuda2001simulated,sharroush2017voltage}

Specifically in the context of MARS, usage of a Zero-$V_{th}$ transistor reduces the oscillator supply voltage to $\sim$0.5V with $\sim$1uA current, allowing the system impedance to match directly with low voltage DC power harvesters (e.g., two photodiodes or the touch of a 3$cm^2$ fingertip  on a  thermoelectric generator) without the need for power management. This convenience reduces the number of components required in the system and the  size of the harvester, thus lowering cost and opening the possibility of making the entire system printable in the future  (\autoref{sec:formfactor}). 

\begin{center}
\begin{figure}[!ht]
\includegraphics[width=15cm]{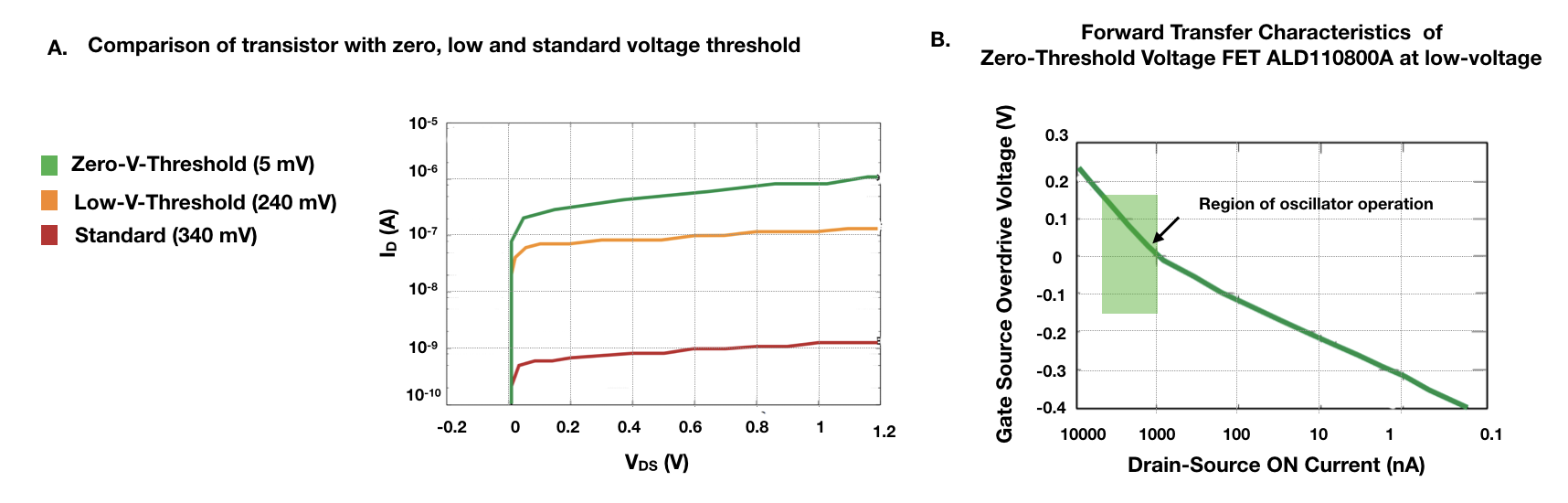}
\caption{\textbf{Theory of Operation : Zero-$V_{th}$ Transistor} \textbf{A.} zero threshold voltage transistors have higher current drive capability in comparison to low threshold and conventional transistors. \cite{junior2012zero} B. Region of operation of zero threshold voltage transistor \cite{ALD110800} exploited in our MARS system } 
\label{fig:theory_of_operation_zero_vth}
\end{figure}
\vspace{-0.25in}
\end{center}

\subsection{MARS Tag Communication}\label{sec: tag_design_comm}
The communication block in the MARS Tag consists of an oscillator, an RF junction field effect transistor that acts as an analog switch, and a antenna. The combination of the ultra-low power oscillator and analog switch results in the nano-watt wireless sensing capability of the MARS tag. 

\subsubsection{\textbf{Low-power and startup voltage Modified Clapp Oscillator (MCO) : Hardware Design and Characterization}} \label{sec:oscillator}

An oscillator is at the core for generating unique frequencies for the FM-backscatter based MARS tags. In this section, we will explore the hardware design and characterization behind MARS's ultra-low power and startup voltage oscillator. The oscillator should satisfy the following design criteria:
\begin{enumerate}
\item Produce a unique frequency with minimal power consumption. 
\item Be sensitive to a large range of input sensor changes to produce a large range of frequency changes.   
\item The $V_{pp}$ of the output sine wave signal from the oscillator should be within a range for the transistor analog switch to perform in the lower triode region to minimize power consumed (explained further in \autoref{subsec:switch}).
\item Low startup-voltage in 100's of mV for the oscillator is preferable as it can allow bypassing the need for buck-boost complex power management and keep the overall number of components for the system low.
\end{enumerate}

We will explain the oscillator hardware design in three stages. First, we start with a conventional common gate LC Colpitts oscillator. Second, we replace its MOSFET with a zero-Vth MOSFET to lower the operation voltage and modify it to the Enhanced Swing Colpitts Oscillator (ESCO) using an additional inductor \cite{machado2015analysis, farhabakhshian2010475}.
Then, we explore the usage of ESCO as a low startup voltage oscillator for FM-backscatter. Finally, we demonstrate our variant of ESCO, a Modified Clapp Oscillator (MCO). We explain its hardware modifications that allow for its startup voltage to be similar to ESCO but also enables lower current consumption (<2 $\mu$W) with similar SNR of the communicated unique frequency.

A conventional Colpitts oscillator (\autoref{fig:theory_of_operation_oscillator_1}A) is a type of LC oscillator consisting of a transistor-based amplifier and a feedback network consisting of an inductor ($L_{1}$) and two capacitors ($C_{1}$, $C_{2}$). The oscillation frequency is determined by the resonant frequency of the LC tank, formed by the inductor ($L_{1}$) connected in parallel to the two capacitors ($C_{1}$, $C_{2}$) connected in series. The frequency can be calculated as $F=\frac{1}{2\pi\sqrt{ L_{1} C_{eq}} }$. 

\begin{center}
\begin{figure}[!ht]
\includegraphics[width=0.9\columnwidth]{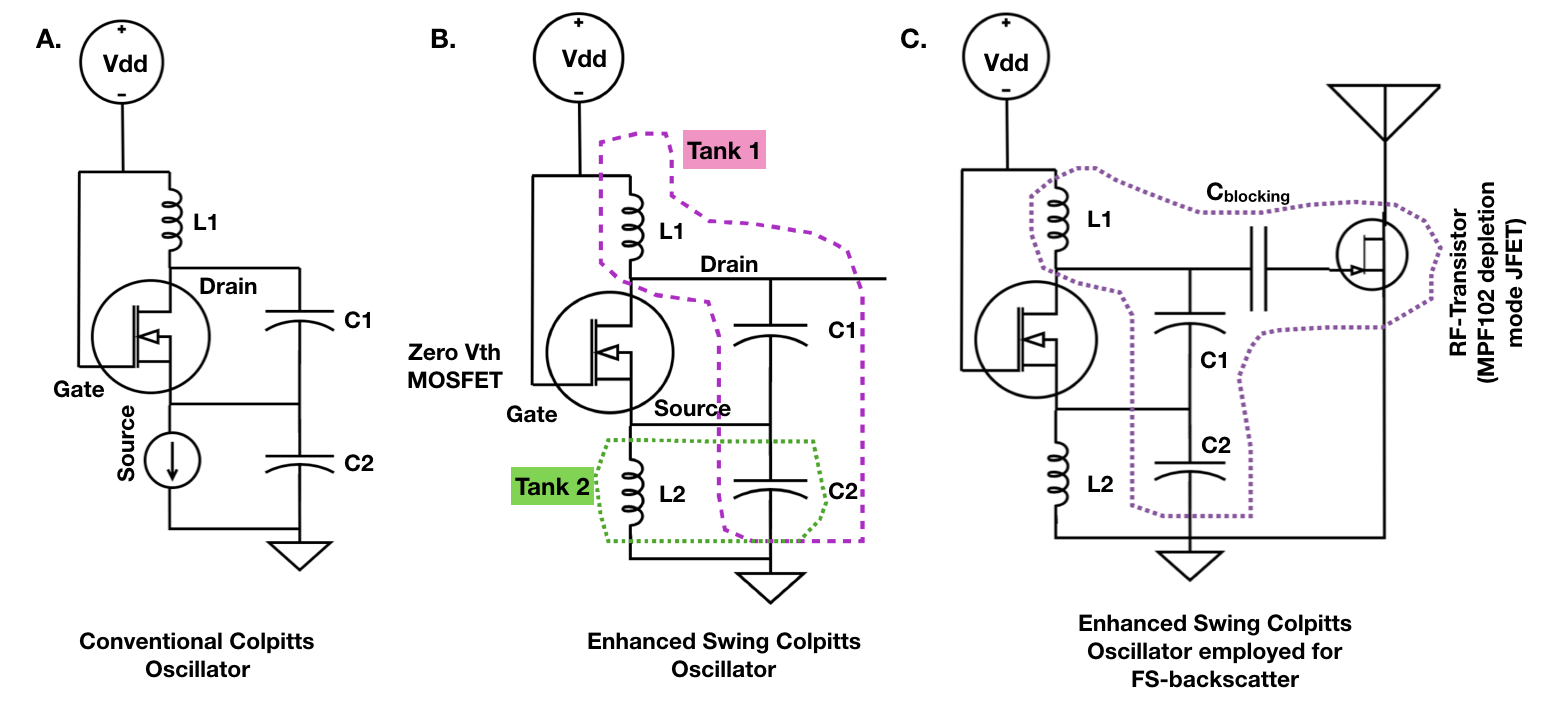}
\caption{\textbf{Steps of Building a Modified Clapp Oscillator (MCO)} \textbf{A.} Start with the conventional Colpitts oscillator configuration \textbf{B.} Replace the MOSFET with a zero-Vth MOSFET and replace the Colpitts current source with a degenerative current source (an inductor) to build the Enhanced Swing Colpitts Oscillator (ESCO) \textbf{C.} Complete ESCO drain-output RF-backscatter circuit }
\label{fig:theory_of_operation_oscillator_1}
\end{figure}
\vspace{-0.2in}
\end{center}

To \textbf{achieve higher amplitude and lower startup voltage, the traditional Colpitts oscillator can be modified from its traditional architecture by changing or adding components.} The \textbf{first component is a zero-Vth MOSFET} (e.g., the ALD110800A \cite{ALD110800}), which can replace the traditional MOSFET to lower the startup voltage of the oscillator as discussed in \autoref{mosfet_primer} \cite{machado2015analysis}.

The second change is to use the Enhanced Swing Colpitts Oscillator (ESCO) configuration \cite{farhabakhshian2010475} (\autoref{fig:theory_of_operation_oscillator_1}B), where \textbf{an inductor, used as a degenerative current source}, is used at the source of the transistor to replace the traditional current source (a resistor) \cite{andreani2005study}. This technique allows us to reduce power consumption, since the inductor only uses reactive power, which can be fed back into the primary tank. Additionally, using an inductor allows our primary tank to swing below zero at the source, thus giving a better output range. Albeit this secondary inductor creates a secondary LC tank with C2, we prevent this secondary tank from affecting our primary tank's oscillation by designing the secondary tank's startup voltage to be much higher than our range of operation.

We tested an ESCO-based backscatter tag tuned at different frequencies ranging from 100kHz-1000kHz (\autoref{table:drain_esco}). The ESCO drain output is fed into $C_{blocking}$ which removes the DC bias before input into the JFET. The JFET changes the impedance connecting the antenna to the ground using the voltage changes at its gate and effectively communicates frequency modulated information  (explained in \autoref{subsec:switch}). Parasitic capacitances from transistors and traces can affect the oscillation frequency. Accounting for parasitic capacitances , we perform a detailed AC analysis of the circuit (\autoref{fig:theory_of_operation_oscillator_1}B in \autoref{Appendix}), which shows the complete LC tank. Using $C_{JFET}$ to denote the effective capacitance formed by the parasitic capacitance network caused by the JFET and the antenna, $C_{JFET}$ and $C_{blocking}$ are connected in series and then connected in parallel with the inductor. Since  $C_{JFET} << C_{blocking}$, where $C_{JFET}$ has a small value in the several to 10s of pF range, it becomes the primary effective capacitance which gets added to the LC tank. Other parasitic capacitances from the zero-Vth MOSFET and traces are also very small (several pF). Thus, they do not affect the oscillation frequency to a large extent. The formula for frequency of the ESCO is the same as the conventional Colpitts. We use $f_{calc}$ to denote calculated frequency using $C_{eff}$ and $L_{1}$ in \autoref{table:drain_esco}.   \autoref{table:drain_esco} shows details of different $f_{measured}$ frequency setups which we built near frequency value f. We powered the ESCO-based MARS tag setup (\autoref{fig:theory_of_operation_oscillator_1}C) using a power supply and report startup voltage ($V_{in}$),  current ($I_{in}$), power (P) and SNR for back-scattering a single unique frequency. We also further calculate the number of photodiodes that would be required to run the backscatter setup  if used in lighting of 500 lux (the final two photodiode system worked well in practice in our laboratory and cafe).

\vspace{0.2in}
\begin{tabular}[b]{ccc ccc ccc ccc cc}\hline
f	& $f_{measured}$ & $f_{calc}$ & $V_{in}$ & Iin & P &  SNR & PD &	L1 & L2 &	C1 & C2  & $C_{block}$ & $C_{eff}$ \cr
            
[kHz]	& [kHz] & [kHz]& [mV] & [uA] & [nW] &  [dB] & &[mH] & [mH] &	[pF] &  [pF] & [nF] & [pF] \cr
\hline 
\textbf{100} & 106 & 105.144 & 270 & 17.05 & 4603 & 15 & 12 & 4.7 & 10 & 1000 & 1000 & 100 & 487.5 \cr
\textbf{200} & 203 & 202.062 & 110 & 3.5 & 385 & 21 & 1 & 4.7&  10&  220&  220& 100& 132\cr
\textbf{300} & 293 & 310.781 & 100 & 2.95& 295& 17 & 1&  4.7&  10& 47& 100& 100& 55.8\cr
\textbf{400} & 414 & 414.953 & 170 & 6.8& 1156& 16& 3&  4.7& 10& 10& 47& 100& 31\cr 
\textbf{500} & 502 & 508.402 & 240 & 11.64 & 2793& 15& 8 & 2&  10 &47& 100& 100& 48.3\cr 
\textbf{600}  & 617&  642.297 & 220 & 11.1& 2442& 22& 6 & 2& 10& 10& 47& 1& 30.7\cr
\textbf{700} & 689 & 672.552 & 200& 9.6& 1920 &27& 5&  1&  1& 47& 94& 100& 54.62\cr
\textbf{800} & 820 & 838.820 & 250& 14.12& 3530& 24& 10& 1& 10 & 20 & 47 & 1 & 36 \cr
\textbf{900} & 925 & 954.840 & 380 & 33.1 & 12578 &20 &30 & 0.3 & 1& 100& 220& 100& 92.61\cr
\textbf{1000} & 999 & 109.0795 & 650 & 143.2& 93080& 27& 220& 0.6& 1 &100& 30& 100& 33.03 \cr
\end{tabular}
\captionof{table}[text for list of tables]{Power consumption and backscatter signal SNR for different ESCO with drain output configuration (PD- Photodiode)}
\label{table:drain_esco}

\begin{center}
\begin{figure}[!ht]
\includegraphics[width=\columnwidth]{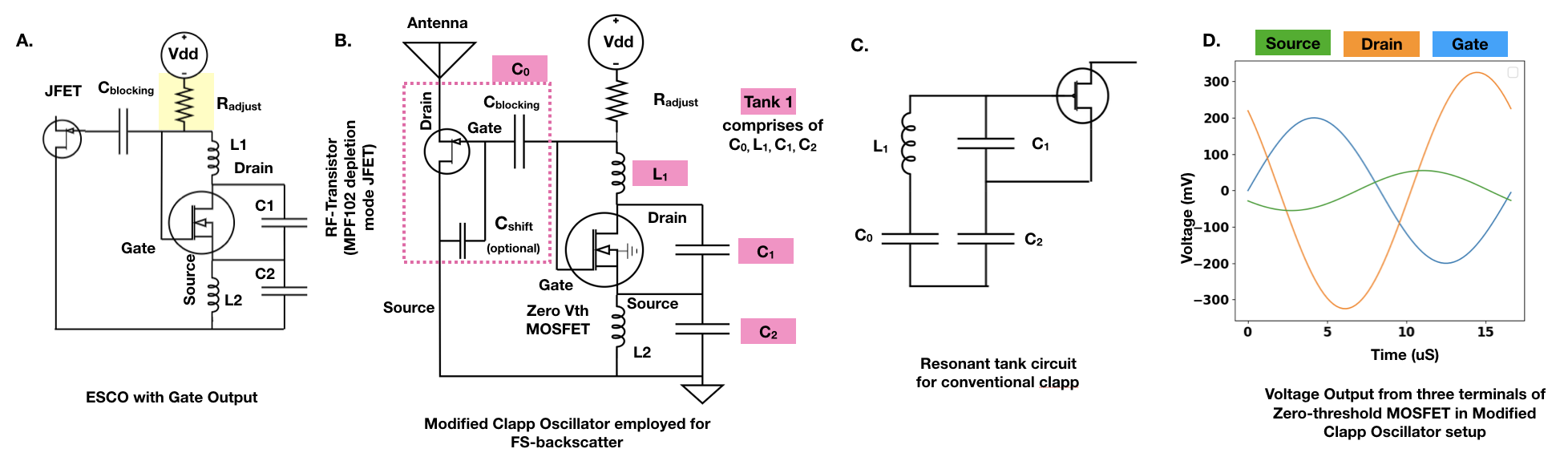}
\caption{\textbf{Steps of Building Modified Clapp Oscillator (Continued)} \textbf{A.} ESCO with gate output with addition of $R_{adjust}$ \textbf{B.} Modified Collpits Oscilator(MCO) with frequency dependent on L1,C1,C2,C0 \textbf{C.} MCO is a type of clapp oscillator \textbf{D.} MCO with all three terminals are oscillating. The gate oscillates to create a negative feedback loop.}
\label{fig:theory_of_operation_oscillator_2}
\end{figure}
\vspace{-0.2in}
\end{center}

While ESCO has a low startup voltage, the current is still on the higher side $>3uA$. Thus, we experimented with the \textbf{addition of a third component, $R_{adjust}$, a current limiting resistor} to prevent the MOSFET from drawing excess current from the power harvester (\autoref{fig:theory_of_operation_oscillator_2}A). This resistor also comes with the side effect of creating an RC tank with the effective capacitance of the primary tank. However, we prevent this from affecting the primary tank's oscillation by making the resistor value high enough to act as an impedance. The oscillations in the primary tank will then limit their flow through the high impedance resistor, and enough current will flow into the lower impedance capacitors to continue oscillations. Unfortunately, this comes with the side effect of increasing the startup voltage for Drain oscillations significantly.

We then changed the circuit a fourth time by \textbf{adding the output to the gate of the Zero-vth MOSFET in the ESCO setup} (\autoref{fig:theory_of_operation_oscillator_2}B). We have found the gate to have a lower startup voltage requirement than the drain. Thus, despite $R_{adjust}$ limiting current and increasing the startup voltage for the drain, we can utilize the gate output with a lower supply voltage. An optional $C_{shift}$ is added to the gate of the JFET to modify the oscillation frequency. A subtle side effect of adding $R_{limit}$ and using the gate as the output is that the effective capacitance, denoted as $C_{0}$, of the capacitor network formed by the parasitic capacitance of JFET ($C_{JFET}$), $C_{blocking}$, and optionally $C_{shift}$ appear as connected in series with the inductor $L_{1}$,  thus changing the circuit into a Clapp oscillator-like configuration (\autoref{fig:theory_of_operation_oscillator_2}C). The detailed configuration of the exact LC tank considering parasitic reactances in the circuit is shown in \autoref{Appendix}. We call this configuration the Modified Clapp Oscillator (MCO). The oscillation frequency is still the resonance frequency of the LC tank. Ignoring other parasitic reactances that has smaller effects to the oscillation, the LC tank is formed by $C_{0}$, $C_{1}$, and $C_{2}$ connected in series and then connected in parallel with $L_{1}$.



Because of the reciprocal rule for adding capacitors connected in series, the small $C_{JFET}$ value dominates $C_{0}$, and $C_{0}$, having smaller value than $C_{1}$ and $C_{2}$, dominates $C_{eq}$, making the oscillation frequency very sensitive to the change of $C_{0}$. As a result, modifying the value for the optional $C_{shift}$, which is in parallel with $C_{JFET}$, can drastically shift the oscillation frequency.

\vspace{0.2in}
\begin{tabular}[b]{ccc ccc ccc ccc ccc}\hline
f	& $f_{meas}$  & $V_{in}$ & $I_{in}$ & P &  SNR & PD & $R_{adj}$ & L1 & L2 & C1 & C2  & $C_{block}$ &  $C_{shift}$\cr
            
[kHz]  & [kHz] & [mV] & [uA] & [nW] &  [dB] &  & KOhm & [mH] & [mH] &	[pF] &  [pF] & [nF] & [pF] \cr

\hline 

\textbf{100} & 103 & 450 & 2.1 & 945 & 15 & 2$^*$  & 200 & 29.4& 1& 100& 220 &100 &30\cr
\textbf{200} & 202 & 450 & 2.02 &909 & 19 & 2$^*$   & 200 & 24.7 & 4.7 & 20 & 100 & 100& n/a \cr
\textbf{300} & 303 & 350 & 1.3& 455& 16& 2$^*$   & 68& 14.7 & 10 & 47 & 47& 100& n/a \cr
\textbf{400} & 389 & 270 & 0.7& 189 &18& 2$^*$  & 82& 10& 10& 20& 47& 100& n/a \cr
\textbf{500} & 502 & 200& 2.15& 430& 18& 1&  56 & 5.7& 1& 20& 100& 100& n/a \cr
\textbf{600} & 602 & 250& 0.5& 125& 15& 1&  84 & 4.7& 10& 10& 10& 1& n/a \cr
\textbf{700} & 686 & 620& 1.4& 868& 17& 2$^*$ & 380& 4.7& 10& 47& 47& 10& 47 \cr
\textbf{800} & 806 & 440& 1.64& 721.6& 15& 2$^*$ & 220& 4.7& 10& 20& 30& 10& 47 \cr
\textbf{900} & 933 & 330& 1.8& 594& 23& 2$^*$  & 39& 2& 4.7& 10& 10& 10& n/a \cr
\textbf{1000} & 1016 & 460& 2.13& 979.8& 17& 2$^*$   & 220& 2& 4.7& 47& 47& 10& 220 \cr
 
\end{tabular}
\captionof{table}[text for list of tables]{Power consumption and signal SNR for different MCO gate output configuration ($^*$connected in series, PD- Photodiode)}
\label{table:gate_clapp}

Note that the behavior of the MCO is different from conventional common gate Colpitts and ESCO where the drain oscillation is an amplified output of the source. In an MCO, all three terminals are oscillating. The inductor $L_{1}$ offers an opposite phase output signal into the gate node, creating negative feedback (\autoref{fig:theory_of_operation_oscillator_2}D). Increasing $R_{adjust}$ leads to an increase in gate oscillation, decrease in drain oscillation, and decrease in overall power consumption. Experimentally, we found that using the gate as output results in a lower start up voltage for the oscillator compared to using the drain as output, making the gate output the better configuration. We found that it is possible to tune $R_{adjust}$ for different frequency setups, so the oscillation output at the gate is around +/- 200mV, which is the ideal operation range for the JFET (green region \autoref{fig:jfet}), where the optimal SNR is produced without generating harmonics.

We performed a similar experiment as we did for the ESCO to measure SNR. As shown in \autoref{table:gate_clapp}, we measured that for all target frequencies from 100kHz to 1MHz, the MCO configuration can be powered from 2 photodiodes under 600 lux, while generating similar SNR compared to the ESCO configuration. For the MCO experiment, $f_{calc}$ is omitted because in practice, the expected oscillation frequency is harder to calculate as the oscillation frequency depends on $C_{JFET}$, which we could not measure. Additionally, our simulation shows an oscillating gate also creates a constantly changing transistor capacitance for the zero-vth MOSFET due to the Miller effect \cite{miller1920dependence}, affecting the oscillation frequency.



In conclusion, leveraging the parasitic reactance of the system, utilizing the gate node as the output, employing $R_{adjust}$ to limit current, and using an inductor as a degenerative current source, we are able to modify the classic Colpitts oscillator into the MCO, which satisfies the four design requirements of a desired oscillator discussed in the beginning of this subsection.

\subsubsection{\textbf{RF-analog Switch}}\label{subsec:switch}
In the next stage, the oscillator output is fed into an RF-switch which is responsible for changing its impedance with changing voltage. To maintain the versatility of the signal being sensed and low-power consumption, we ideally want the switch to satisfy the  following design parameters:
\begin{enumerate}
\item The switch should be able to have linear impedance changes with voltage input to be able to communicate different types of sensed information. 

\item Preferably not require constant $V_{dd}$ for operation to lower the power consumption. This preference reduces requirements for power management to handle voltage mismatch between the power harvester and the RF-switch. Additionally, it also reduces quiescent current wastage.

\item For a given signal to create impedance changes, the corresponding current consumption should be as low as possible.

\end{enumerate}

\begin{center}
\begin{figure}[!ht]
\includegraphics[width=0.6\columnwidth]{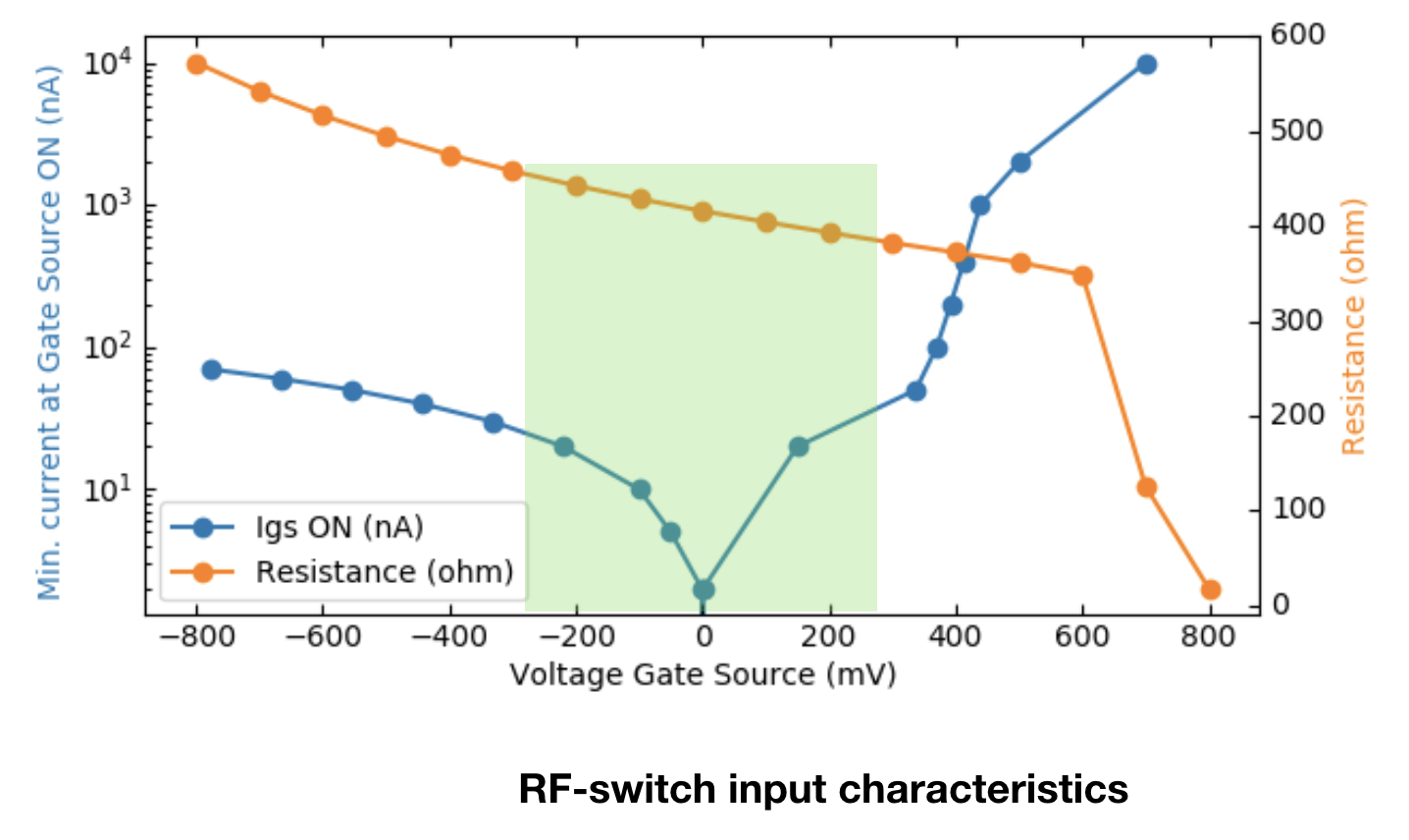}
\caption{\textbf{Characteristics of the RF-switch JFET}: Voltage versus current and resistance graph. Region marked in green is the ideal  operation region for JFET where the power consumption is low and impedance changes are linear.}
\label{fig:jfet}
\end{figure}
\vspace{-0.2in}
\end{center}

In our system,  we use a JFET  (MPF-102) as an RF switch in the common-source configuration in the MARS tag (\autoref{fig:theory_of_operation_oscillator_2}b). It does not require a constant $V_{dd}$ like the commonly used ADG-902 RF switch IC in similar works \cite{ubiquitouch,ranganathan2018rf}.
\autoref{fig:jfet} shows the relationship between $I_{GS}$, $V_{GS}$ and $R_{DS}$ of the JFET measured using an Agilent E5272A. The power consumption for the impedance modulation is ideal for a low power system. As the gate output of the MCO goes through $C_{blocking}$ and inputs into the JFET an AC signal with a range of $\sim$ +/- 200mV  (green region in \autoref{fig:jfet}), the current consumed by JFET is 100's of nA, which keeps the total power consumed low. In addition, in this voltage range, the JFET operates in the ohmic/ triode region (before $V_{pinchoff}$ = -1.9V) and thus will have a desirable linear change in resistance (orange) in response to the sine wave-like signal from the MCO, resulting in few harmonics in the frequency shifted backscatter signal.

\subsubsection{\textbf{Antenna}}
The carrier frequency used by the transmitter and the receiver determines the frequency of antenna used (e.g., 915MHz) in our current setup. We further performed antenna selection based on the trade-offs between antenna gain, form factor, and commercial availability. We use a flexible linear polarization patch antenna (Taoglas FXP290.07.0100A) tuned to the UHF band with 1.5 dBi gain, resulting in a 3.5x4.5x0.1 $cm^3$ size in the MARS tag. While the uniformly polarized commercial whip antenna with 4 dBi gain is more robust to directionality and has higher signal-to-noise ratio (SNR), it is too  thick and big (16x0.5 x1.2 $cm^3$) to be included in MARS tags. 
To optimise antenna size further, in the future, custom printed 915 MHz dipole or PIFA antenna can be included in the prototype, which would be much smaller in area than the current commercial one in use. Another alternative is to use commercially available chip antennas (e.g., the ACAG1204-915-T) with a reasonably high gain of 3.2 dbi and a smaller form factor 1.2 x 0.4 x 0.16 $cm^3$. 

\subsection{\textbf{MARS Tag Analog Sensing}} 
\label{Sensing}

The MARS communication block interfaces with multiple analog sensors by modifying different electrical properties of C1, C2, and L1 that control the oscillator's frequency. We use C0 to tune the oscillator frequency but not for sensing.
\autoref{fig:analog_sensing}A shows positions of inductor and capacitor passives of the LC oscillator in blue, which can be modified directly by placing an analog sensor in its place or indirectly by placing it in parallel or series. Below we will discuss three different approaches to add sensing to the MARS tags, using inductance, capacitance, or self-powered voltage generating  sensor to control the tag's oscillation.

\begin{center}
\vspace{-0.15in}
\begin{figure}[!ht]
\includegraphics[width=\columnwidth]{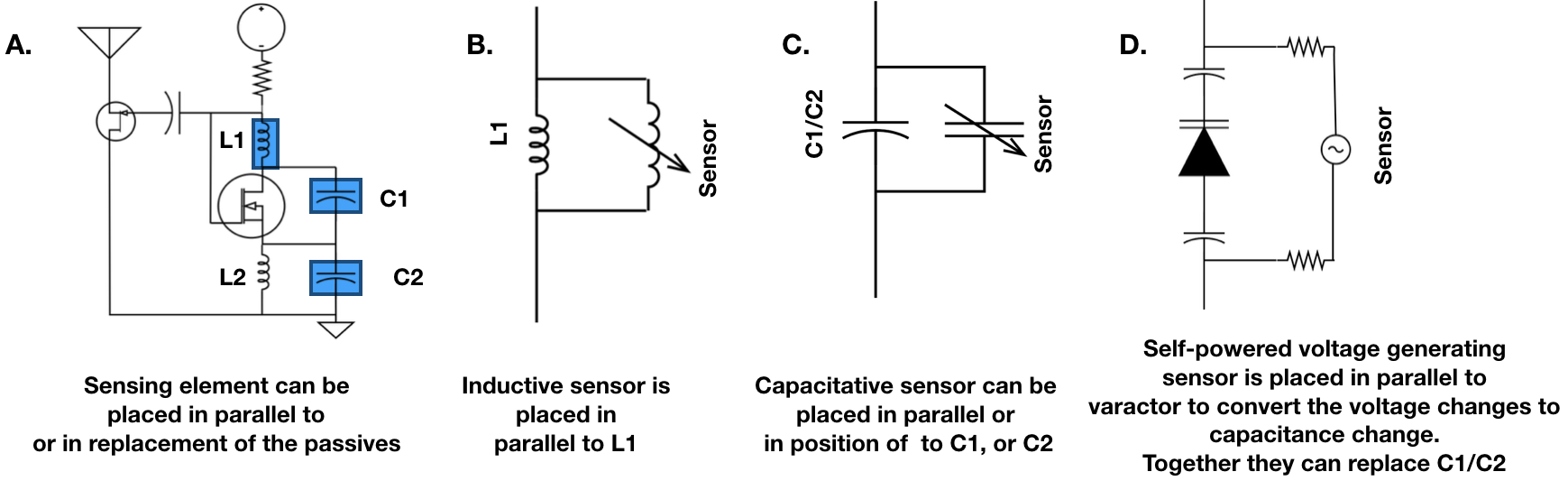}
\caption{ \textbf{Overview of Analog Sensing Modalities:} \textbf{A.} Analog sensors can be placed in parallel with the highlighted components, or can replace the highlighted components entirely. \textbf{B.} Inductive based sensors can be added in parallel with L1 \textbf{C.} Capacitive sensors can be placed in parallel with or replace C1 or C2. \textbf{D.} Sensors which generate voltage can be placed in parallel to a varactor which converts the voltage changes to capacitance changes. This combination can be used to replace C1/C2 in the oscillator.} 
\label{fig:analog_sensing}
\end{figure}
\vspace{-0.3in}
\end{center}

\subsubsection{\textbf{Inductor-controlled oscillator:}} 
We can place an inductive analog sensor in parallel/series with the inductive passive (L1). Thus, any change in the inductance value of the sensor would also change the effective L1 of the oscillator system, effectively changing the frequency of the oscillator (\autoref{fig:analog_sensing}B). We leverage this phenomenon to develop an inductive multi-button-touch game controller (\autoref{fig:multi_touch}).
    
\subsubsection{\textbf{Capacitor-controlled oscillator:}} We can use an analog variable capacitance-based sensor to modify the effective capacitance in the LC tank of the oscillator (\autoref{fig:analog_sensing}C). The sensitivity and frequency modulation range (channel width) of the system can be tuned by adding the sensor into the capacitor network ($C_{0}$, $C_{1}$, and $C_{2}$) at different nodes.
The low capacitance of both C1 and C2 (in pF) gives an extensive range of capacitance sensors. We use the change in pF of capacitance for a swipe-based touch sensor for sensing direction and ID (\autoref{fig:direction} and \autoref{fig:id}).

\subsubsection{\textbf{Voltage-controlled oscillator:}} Several self-powered sensors produce voltage changes from the phenomenon they sense (e.g., a photodiode from light, a piezoelectric or triboelectric generator from mechanical vibrations). Voltage from such self-powered sensors can be fed into a variable capacitor diode to produce corresponding capacitor changes (\autoref{fig:analog_sensing}D). When tuned to the proper range, this new combined self-powered sensor and variable capacitor diode can replace either capacitor (C1/C2) in the oscillator circuit. We have leveraged this setup to communicate speech using the MARS tag (\autoref{fig:speech}).

\subsection{MARS Tag Power Harvesting}\label{power_harvesting}
For the MARS system, both the startup supply voltage ($\sim$ 500 $mV$) and the current ($\sim$ 2 $\mu{A}$) are relatively low for different oscillator setups (detailed later in  \autoref{table:gate_clapp}) than conventional systems. This low supply voltage and current allows for different DC power harvesting methods, such as ambient light and body heat, to power the MARS tags, eliminating the need for complex circuitry or power management. Below, we discuss the two power harvesting techniques we have so far explored.

\vspace{-0.1in}
\begin{center}
\begin{figure}[!ht]
\includegraphics[width=\columnwidth]{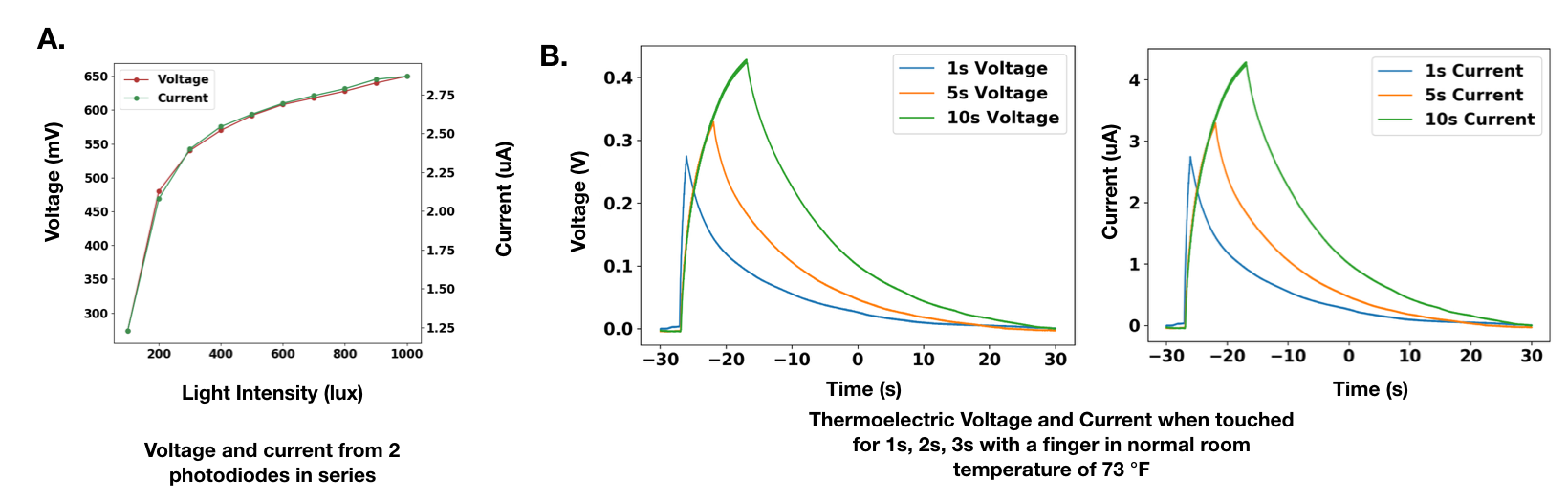}
\caption{\textbf{Power Harvester Characterizations:} \textbf{A.} The voltage and current produced by two photodiodes in a series electrical configuration. \textbf{B.} The voltage and current produced by a thermoelectric cell pressed for 1, 2, and 3 seconds in a room at $73^{\circ}F$.} 
\label{fig:power_characterisation}
\end{figure}
\end{center}  
\vspace{-0.3in}

\subsubsection{\textbf{Photodiode:}} Lighting in a general office environment is about 500-1000 lux. \autoref{fig:power_characterisation}A demonstrates the voltage and current characterization of two photodiodes in series, which we use to power a MARS tag reliably in different frequency configurations (\autoref{table:gate_clapp}). We change the light intensity of the incandescent light bulb up to 60 $W$ and in series measure light intensity with a lux meter for these experiments. The voltage and current readings were done across 56 Kohm resistance, similar to $R_{adjust}$.

\subsubsection{\textbf{Thermo-electric Generator:}}  
Thermoelectric power generation varies with many factors (e.g., room temperature, body temp, area of contact, the pressure of contact). We conservatively reported (\autoref{fig:power_characterisation}B) voltage and current from the thermoelectric generator (Perpetua Technologies) across a 100 Kohm resistor when an adult's thumb with a body temperature of 97.8 °F touches for 1s, 5s, and 10s in a room at 73°F. The peaks of 1s touch curve at  0.4V and 2.5uA demonstrate that the power generated is sufficient to power  a  MARS tag (\autoref{table:gate_clapp}) with the touch of a human finger. 

In conclusion, leveraging a Modified Clapp Oscillator (MCO) (\autoref{sec:oscillator}), we have been able to achieve analog sensing of inductance, capacitance and voltage, in $<1uW$ with simple circuitry (2 active and 7 passive components). Low startup voltage and low-power allows MARS tags to be powered by 2 photodiodes in indoor ambient light or the touch of a finger on a thermogenerator.

\section{Interaction-specific MARS TAG Design and Applications }
\label{SpecificTagDesign}
In this section, we discuss interaction-specific circuit design changes in MARS tags. We start by detailing our experimental setup, and then we explain the common part of the transceiver pipeline, which all interactions share. Next, we provide the details of several interactions which MARS tags can support (\autoref{fig:prototype_application_summary}). For each interaction we demonstrate a built prototype, the real-time signal we record by interacting with it, the transceiver-specific pipeline for processing it, and finally, the applications MARS tags can support. 
\begin{table}[!ht]
  \includegraphics[width=\columnwidth]{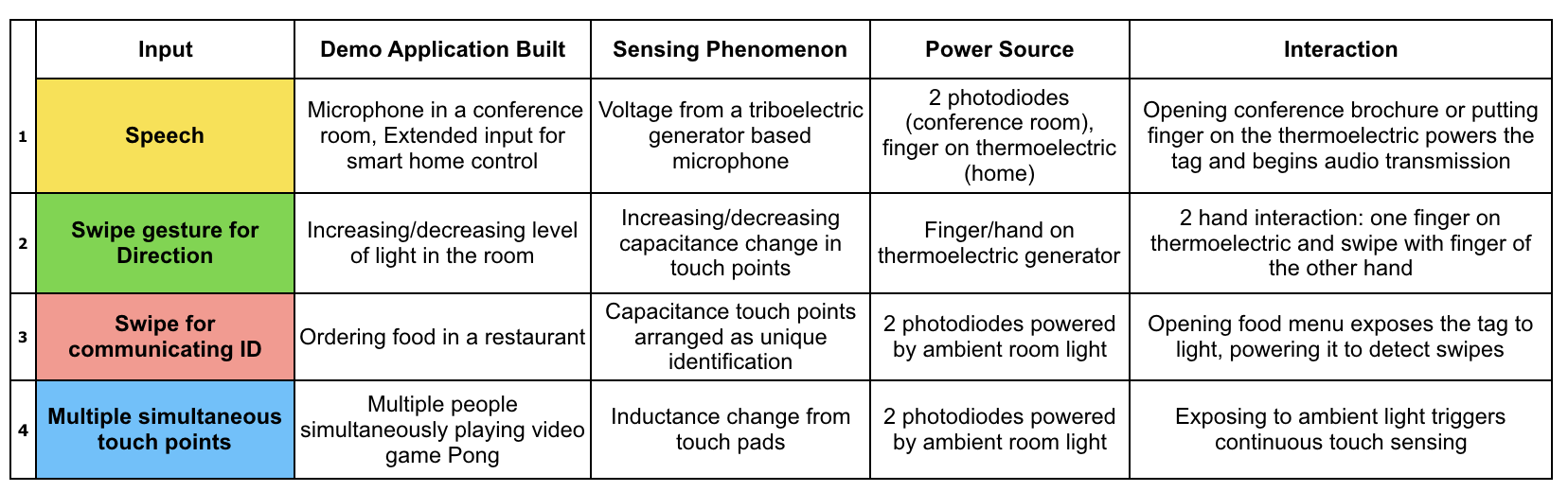}
  \caption{\textbf{Prototype Applications:} Summary of input gestures that can be detected using MARS and corresponding details of the prototype and the demo application.}
  \label{fig:prototype_application_summary}
\vspace{-0.3in}
\end{table}

\subsection{Experimental Setup}
Our system consists of three main components -- the transmitter (TX), MARS tags, and the receiver (RX). We use an Ettus Research N210 USRP with a UBX-40 USRP daughterboard as a transceiver in a monostatic backscatter configuration with a separate antenna for transmitting and receiving. For our experiments, the USRP together Nooelec LaNA amplifier produces a 915 $MHz$ carrier wave (16 dBm) at 29 dBm (within FCC limit) \cite{FCC}. The transmitter and receiver antenna (circularly polarized 6 dBi) were placed on the ceiling of the room, approximately 7 feet from the MARS tags.

\subsection{Transceiver Processing}\label{sec:processing}
\begin{center}
\begin{figure}[!ht]
\includegraphics[width=\columnwidth]{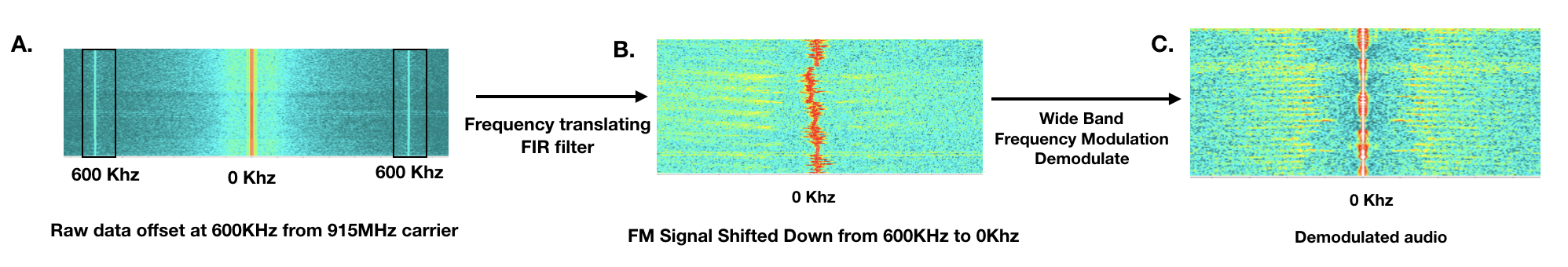}
\caption{\textbf{FM Transceiver Pipeline for Audio:} The three signal processing steps displayed here create the pipeline which is used for: \textbf{A.} receiving the backscattered signals \textbf{B.} shifting backscatter signals to 0 Hz where they can then be \textbf{C.} demodulated into human perceptible audio data} 
\label{fig:processing}
\end{figure}
\vspace{-0.3in}
\end{center} 
We leverage the GNU Radio Companion (GRC) software suite to create a processing pipeline for each of our four interactions. Each GRC flow graph is optimized for a particular interaction modality, but they all share a similar set of initial digital signal processing steps. We will explain them with an example of audio data (\autoref{fig:processing}). First, the data is collected at $1e^6$ Hz and a low pass filter is applied (\autoref{fig:processing}b). Second, a frequency translation step is applied to shift the backscattered signal down to 0 $Hz$ offset, e.g., from 600 $kHz$ to 0 $Hz$ (\autoref{fig:processing}b). Finally, for frequency modulated (FM) signals like audio, we utilize a Wide Band Frequency Modulation (WBFM) demodulation block to separate the backscattered signal from the backscattered carrier wave (\autoref{fig:processing}c).

\subsection{Speech}
\begin{center}
\begin{figure}[!ht]
\includegraphics[width=\columnwidth]{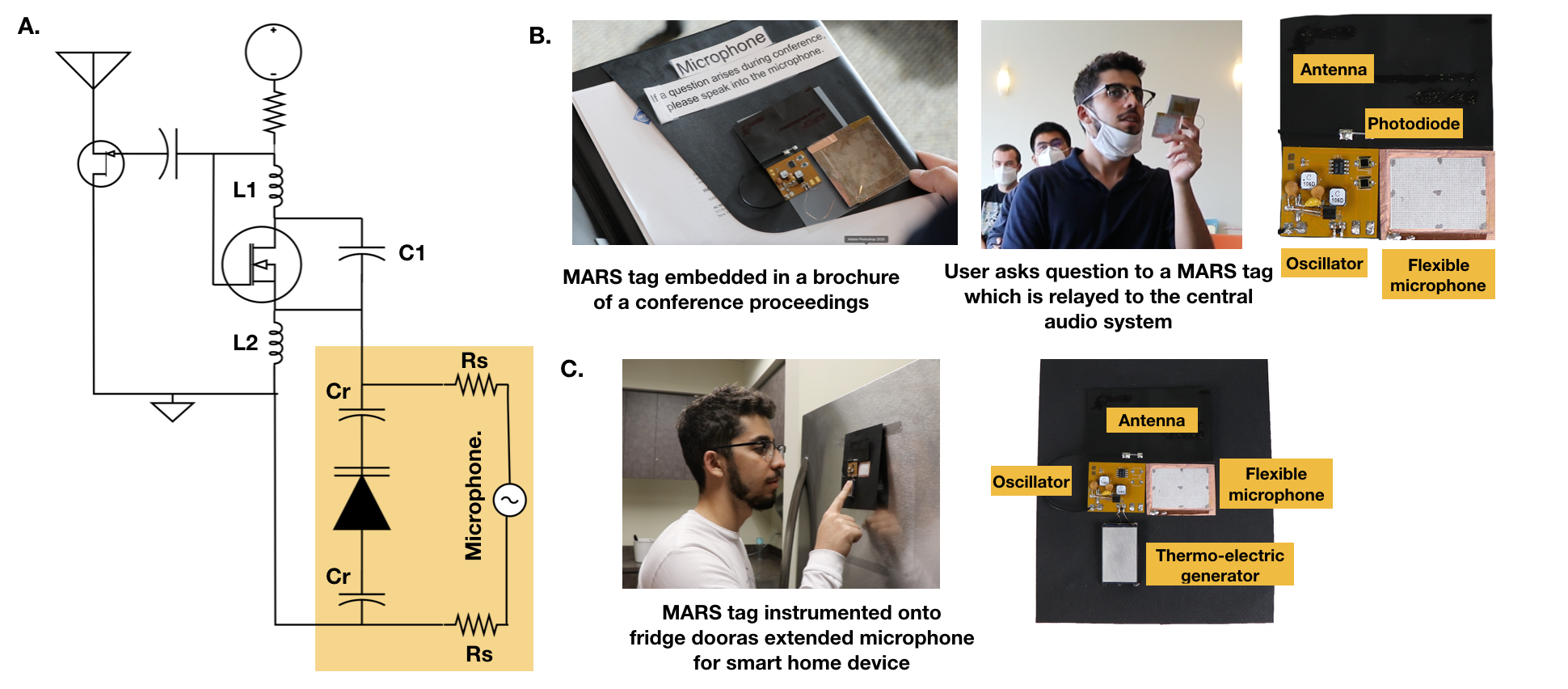}
\caption{\textbf{MARS Facilitated Wireless Speech Transmission: }\textbf{A.} Audio signals from the microphone are utilized to change the capacitance of a varactor which modulates the oscillation frequency of the backscattered signal. \textbf{B.} Conference program insert MARS wireless microphone, demonstration of the MARS wireless microphone in a conference environment, and close up of the powered prototype \textbf{ C.} MARS wireless microphone attached to a foam pad for affixing to home appliances. Demonstration of utilizing the MARS wireless microphone, powered by a touch of a finger, to extend the range of a smart home device.} 
\label{fig:speech}
\end{figure}
\vspace{-0.3in}
\end{center} 


We created a MARS tag which supports the wireless communication of audio data using our previous work SATURN, a self-powered flexible microphone based on the principle of triboelectric nanogeneration (TENG) \cite{arora2018saturn} as shown in circuit in \autoref{fig:speech}A,C. The microphone is placed in parallel to a varactor (SMV1702-011LF), which changes the capacitance across its ends when the user speaks into the microphone. We also add extra blocking caps Cr, so the microphone's charge input is used just for changing the capacitance of the varactor and does not affect the oscillation in the circuit. This microphone-varactor combination could replace either C1 or C2, but since C2 gives less frequency shift compared to C1, we put it across C2, where it was more stable.  The processing pipeline followed for audio was mentioned before in \autoref{sec:processing}. Audio in the tag takes a +/- 60kHz frequency modulated signal. \autoref{fig:speech} B and C show example applications of MARS audio communication. \autoref{fig:speech}B is powered by two photodiodes, while \autoref{fig:speech}C is powered by a thermometric generator, as mentioned in \autoref{power_harvesting}. Since both the MARS tag and the microphone are thin, we envision that the MARS audio patches can be embedded in a conference brochure and used as an extended microphone for central audio control. Another application is at home, where users can touch and talk with the MARS tag, which acts as an extended microphone for smart home devices. We characterise the quality of audio with increasing distance based on PESQ score in \autoref{sec:sys_characterisation}.

\subsection{Swipe based direction sensing }
MARS can enable swipe-based direction control, leveraging capacitance sensing. The circuit exploits that the human body acts similarly to a bag of water, and any touch interaction over a conductive surface directly or indirectly changes its capacitance.  \autoref{fig:direction}A shows the prototype of a light dimmer, where the user increases or decreases light intensity by swiping a finger over capacitive pads (\autoref{fig:direction}C) while the other finger remains on the thermoelectric to power the circuit from body heat. \autoref{fig:direction}B shows the controller's circuit diagram and capacitative swiping interface, highlighted in green as a variable capacitor at the terminal between C1 and C2. 

\begin{center}
\begin{figure}[!ht]
\includegraphics[width=\columnwidth]{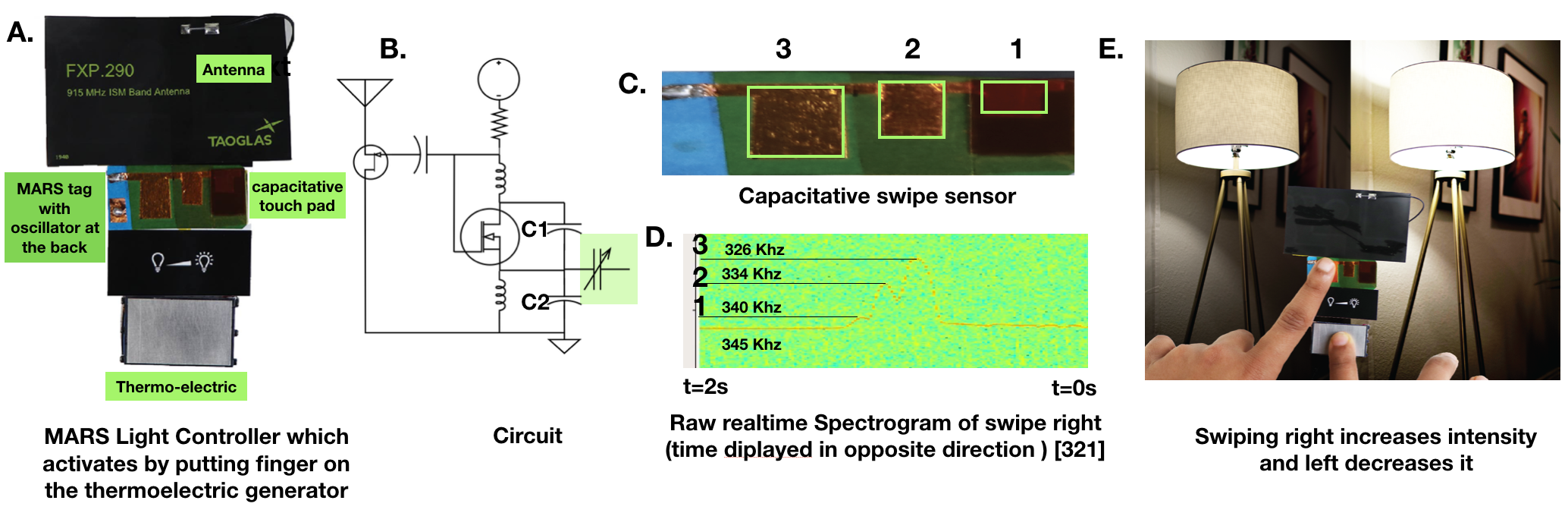}
\caption{\textbf{MARS Facilitated Capacitive Dimmer Demo:} {A.} MARS swipe direction sensor device prototype. {B.}The capacitive touchpad acts as a variable capacitor to modify the oscillating frequency of the MARS tag. {C.} The touchpad sensor is constructed out of copper tape and adhesive backed polyamide sheet (Kapton). Various size pads create varying amounts of electrical capacitance. {D.} Three peaks from the touchpad are detectable on the GRC based transceiver's spectrogram {E.} The swipe output of the MARS tag was interfaced with a Phillips Hue light to act as a dimmer. } 
\label{fig:direction}
\end{figure}
\vspace{-0.3in}
\end{center}

We developed a swiping surface that monotonously increases or decreases its  capacitance depending on the direction of swiping. It is built using paper, copper foil (conductive surface) and kapton tape of two different thicknesses. First, teeth of different sizes were cut from the copper foil and pasted onto the paper. It is followed by layering kapton tape individually on each tooth individually to create different capacitances. For example, digit 1 has three layers of Kapton sheets and will have the least amount of capacitance change when someone touches it while digit 3 has one layer and thus has the maximum capacitance change when someone touches it. We performed a short study with 5 participants, to confirm that each of the digits fabricated are differentiated enough in their capacitance change when a human finger touches it (\autoref{fig:direction_processing}A). We seldom observe borderline cases, and in the future capacitative sensors can be further tuned  by playing with the surface area of the tooth, number of layers of Kapton/or other materials on top of it.

After developing the capacitative touch sensor, and testing its capacitative changes, we next integrate it with the MARS tag. \autoref{fig:direction}D shows the raw spectrogram  from GNUradio of a swipe right. We noticed that a small 2pF change in our circuit produces approximately 5 kHz of frequency change. Thus, we can see in \autoref{fig:direction}d that a swipe right changed a tag tuned to 345 kHz, first to 326 kHz (19 kHz shift)), 334 kHz (11 $kHz$ shift), and 340 kHz (5 kHz shift), respectively. We passed this data through our offline python processing pipeline of Fourier transform (FFT), argmax, smoothing (\autoref{fig:direction_processing}) and linear regression. The slope of this linear regression corresponds to the direction of the swipe. In the future, we will evaluate building a real-time direction sensing pipeline using this strategy.  

\begin{center}
\begin{figure}[!ht]
\includegraphics[width=\columnwidth]{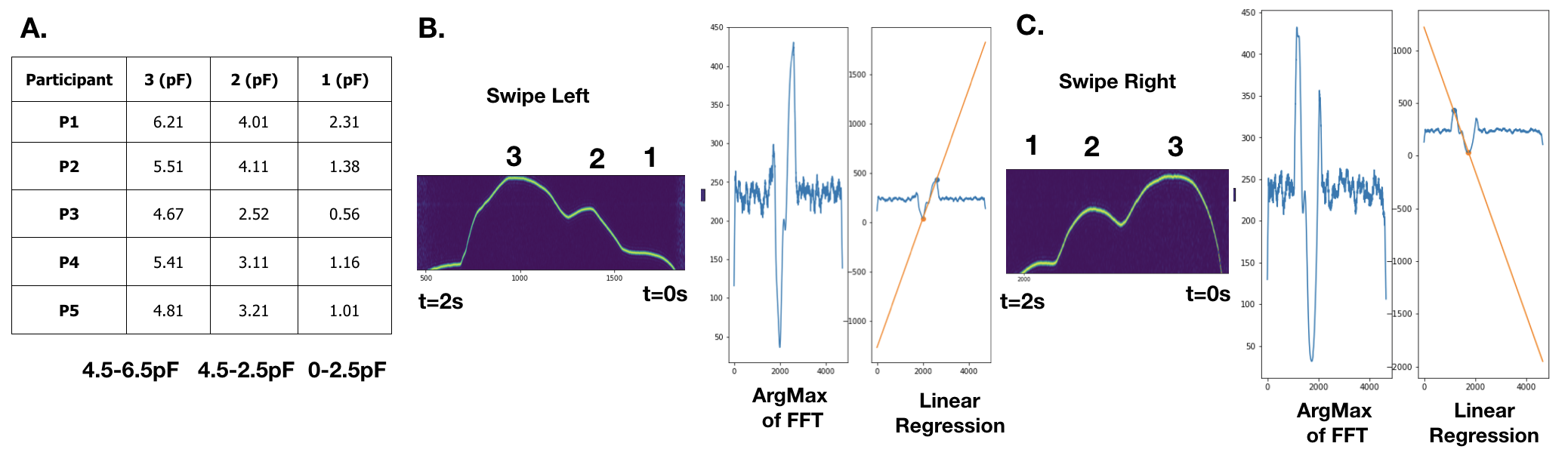}
\caption{\textbf{Capacitive Swipe Sensor Processing:} \textbf{A.} The capacitance of each slider pad varies slightly between participants. \textbf{B.} The signal processing steps for a left swipe: the spectrogram is generated, the spectrogram is filtered into a 1D time-series, a linear regression is found between the maximum and minimum points of the signal. \textbf{C.} The same signal processing pipeline is demonstrated for a right swipe. Notably, the regression line's slope is opposite of the left swipe.} 
\label{fig:direction_processing}
\end{figure}
\vspace{-0.3in}
\end{center}

\subsection{Swipe based Unique ID}
\begin{center}
\begin{figure}[!ht]
\includegraphics[width=\columnwidth]{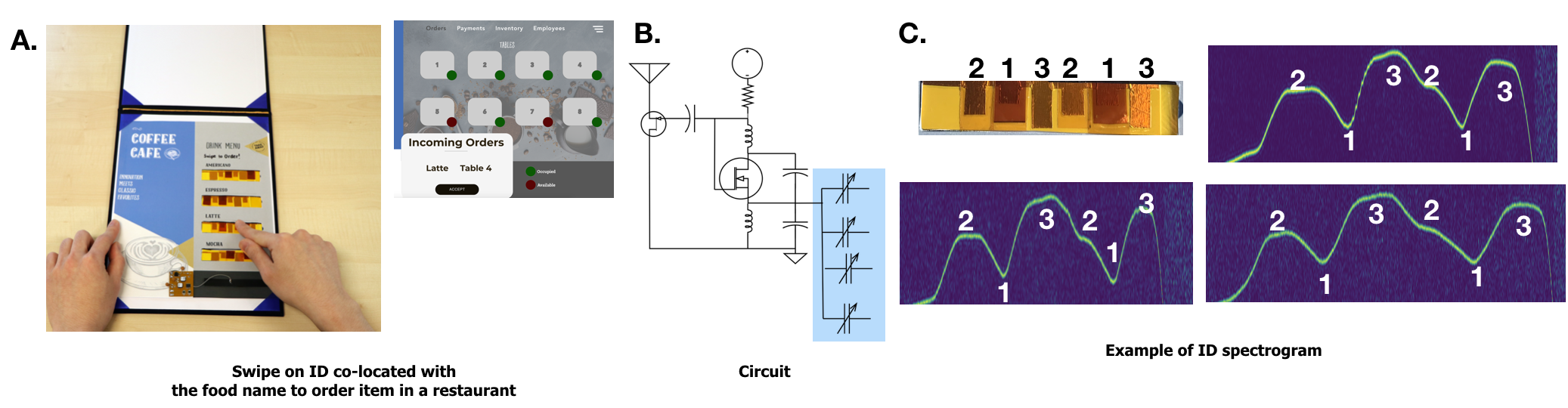}
\caption{\textbf{MARS Facilitated Capacitive Menu Demo:} \textbf{A.} The menu contains a MARS tag and a corresponding capacitive ID sensor for each menu item. When the user swipes an ID sensor on the menu, the signal is transmitted to a remote application for servers to manage the orders. \textbf{B.} The ID sensors act as variable capacitors which modify the oscillating frequency of the MARS tag when they are touched. \textbf{C.} Pictured is an ID sensor with its capacitive pad values labeled along with corresponding spectrograms which illustrate the frequency changes in backscatter signal produced by touching the ID sensor.} 
\label{fig:id}
\end{figure}
\vspace{-0.3in}
\end{center} 
We extend the idea of a capacitative swipe sensor to transmitting ID information by just a finger swipe over six capacitive teeth in a barcode. Adding these barcode IDs to MARS expands the interaction space for one particular channel, opening possibilities for new applications. \autoref{fig:id}A shows a food menu with MARS tag, which has four IDs co-located with each food item. As the user opens the menu, the photo-diode receives light and starts operating the MARS tag, indicating table occupancy. When the user wants to order something, they can do so by swiping on the ID next to the food item, which changes the MARS tag's frequency according to the physical ID swiped. Each restaurant table can have a menu with a MARS tag tuned to a particular frequency. 

Each of the four IDs is a variable capacitor and is attached to our original MARS circuit, as shown in \autoref{fig:id}B. \autoref{fig:id}C shows a close-up view of one of the IDs and the spectrograms' changes when the user swipes it. Over multiple finger-swipe runs on different IDs from the menu, and with MARS tag tuned at 345 $kHz$, we obtained peaks between 318-325 $kHz$ for digit 3, 330-335 $kHz$ for digit 2, and  339-341 $kHz$ for digit 1. In the future, we will employ template matching algorithms like dynamic time warping (DTW) to create an online ID detection pipeline and real-time deployment of different battery-free food menus for the restaurant scenario.

\subsection{Multiple Discrete Touch points}\label{sec: touch}
 
MARS tags can detect multiple discrete touchpoints simultaneously, leveraging inductance-based touch sensing. \autoref{fig:multi_touch}A shows a paper game controller with two buttons:  up and down arrow keys. A MARS tag is attached to it, and it is powered by two photo-diodes. \autoref{fig:multi_touch}B shows the circuit of the game controller. L11 and L12 are two inductors placed below the arrow key buttons, which act as a switch in the circuit. Initially, when no game is being played, L11 is in series with L12, so the effective $L1_{eq}$ = L1|| (L11+L12) in steady-state. When any one of the buttons is pressed (e.g., the up arrow key), the inductor is parallel to it, and L1 is effectively shorted. This shorting makes the inductance $L1_{eq}$ = L1|| L12. Similarly, when down is pressed, then the inductance becomes $L1_{eq}$ = L1|| L11. Each change to $L1_{eq}$ results in a change in frequency. For example, when the up arrow is pressed for a button controller tuned at 289 $kHz$, the frequency shifts to 303 $kHz$ (+14 $kHz$). The down arrow shifts the frequency to 314 $kHz$ (+25 $kHz$) (\autoref{fig:multi_touch}C). For another controller tuned at 349 $kHz$, it shifts to 366 $kHz$ for the up button and 379 $kHz$ for down. The shift can be tuned further to allow packing together a single controller with more buttons or more controllers in a space. A simple threshold based algorithm is used for detection of up/down keys for controlling the game Pong (\autoref{fig:multi_touch}C).  


\begin{center}
\begin{figure}[!ht]
\includegraphics[width=\columnwidth]{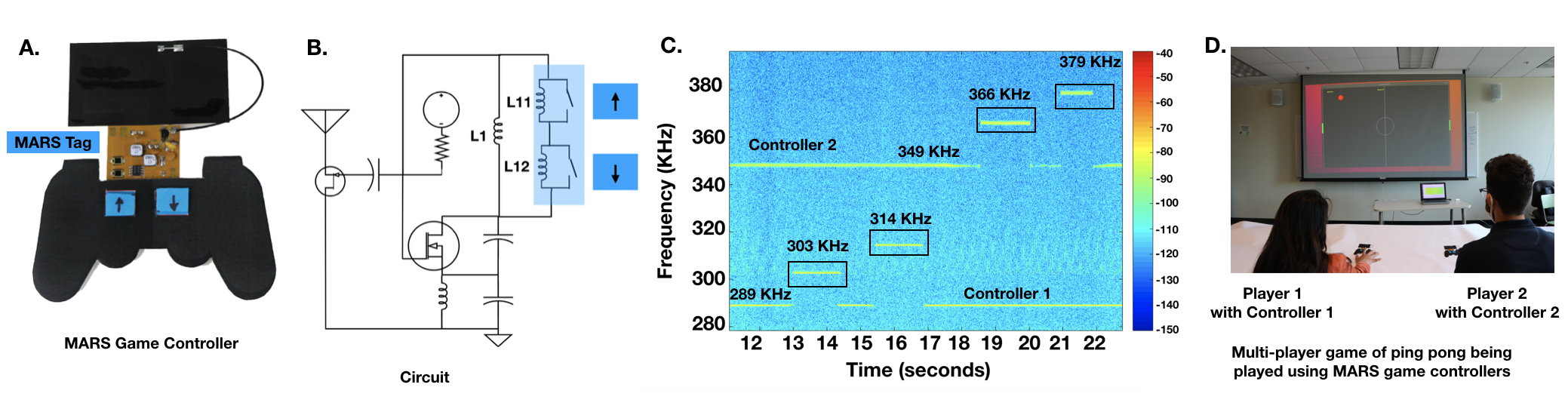}
\caption{\textbf{Pong Game Controller Demo:} \textbf{A.} Paper game controller attached to a flexible MARS tag. \textbf{B.} Touch sensitive buttons switch the MARS oscillator's frequency by shorting a pair of inductors. \textbf{C.} The spectrogram shows the two controller carrier signals (at 289kHz and 349kHz) and their corresponding button press frequency shifts (303 and 314kHz for controller one and 366 and 379kHz for controller two). \textbf{D.} The backscatter button signals are processed in GRC and are fed as control inputs to Pong.} 
\label{fig:multi_touch}
\end{figure}
\vspace{-0.3in}
\end{center}

\section{System Level Performance Characterizations} \label{sec:sys_characterisation}

\subsection{Operational Range}\label{range_experiments}
Backscatter communication is governed by the Friss path loss equation; the power of the radio wave reduces with distance. Thus, the closer the tag is to the transmitter, the stronger the reflected power. For range experiments, the transmitter and receiver antennas were placed next to each other in a monostatic configuration and in line of sight with the tag. The MARS tag was moved away from the antennas incrementally by 2 feet, and both SNR and audio data was recorded. A MARS tag tuned to 350 $kHz$ offset was chosen and powered by a regulated DC power source of $<1\mu {W}$ to ensure reproducible measurements with only the minimal power necessary to start the oscillator.

To calculate SNR, the recorded data was filtered using a moving average and plotted in a spectrogram.  \autoref{fig:range} shows changes in the SNR with distance. The maximum SNR was measured as 45 dB at a distance of three feet. The steepest drop in SNR occurs between the distances of three and nine feet. At a distance of nine feet, the SNR is 38 dB. At the longest distance tested (49 feet) the SNR was measured as 15 dB. 

For calculating audio quality, we used a MARS tag with a flexible microphone placed next to a speaker playing an audio file. The selected audio file was a composition of spoken words commonly used in telecommunication audio experiments.  To mimic the average intensity of human speech sound, we configured the speaker to output in a range of 70-80 dBL. After the data was recorded through the MARS system, the PESQ score, a commonly used measure of quality of audio in telephony systems, was utilized \cite{hu2007evaluation}. A PESQ audio score of four is maximum and one is just understandable to human ears. After some basic background noise filtering, our audio samples achieved a PESQ score of 2.45 at 3 feet. The score drops to 2 at approximately 9 feet and further reduces to 1.75  at 19 feet, and 1.25 at 30 feet.

\begin{center}
\begin{figure}[!ht]
\includegraphics[width=0.9\columnwidth]{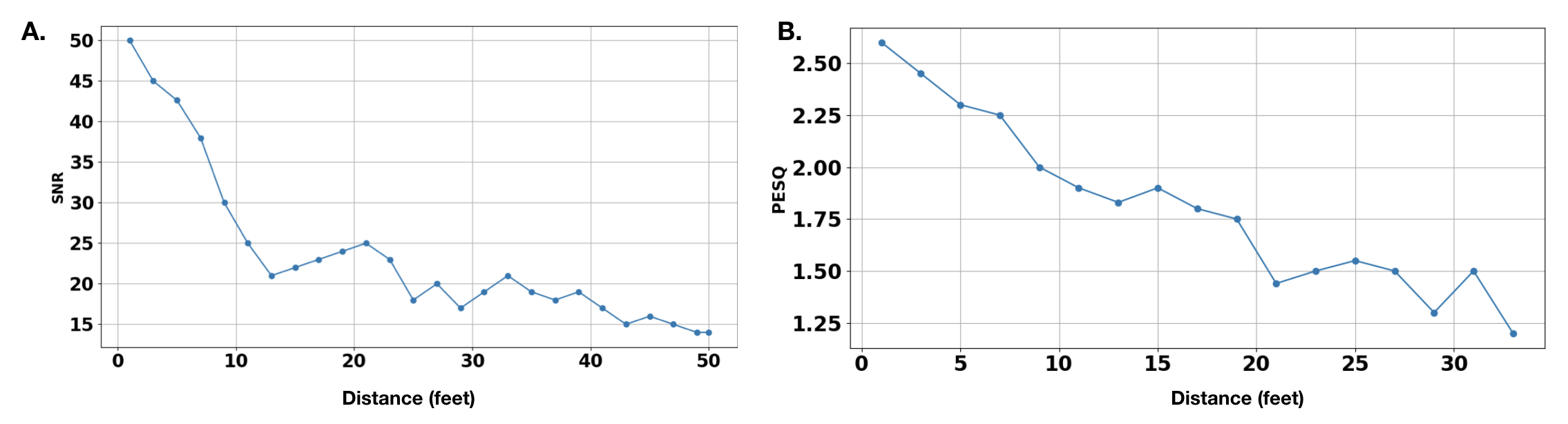}
\caption{\textbf{System Level Characterization}: a) SNR versus distance b) Audio quality score versus distance } 
\label{fig:range}
\end{figure}
\vspace{-0.3in}
\end{center}

\subsection{Multiple tags}
Our experiments demonstrate that a bandwidth of 20 $kHz$ is sufficient for detecting the state of 4 touch points, single swipe direction, or single ID. This result suggests that 30 of such tags can be placed within a sensing range of \textit{40 ft/12m (monostatic)}. An audio transmission scoring PESQ 2 or higher requires +/- 60 $kHz$ bandwidth (120 $kHz$ total bandwidth). Even under the maximum bandwidth of +/-60 $kHz$ for audio, our system can support $\sim$ 8 tags (1 $MHz$ / 120 $kHz$) within the \textit{30ft/9m area of sensing (monostatic)}. Detecting simultaneous tags is computationally intensive; the efficiency of the receiver processing pipeline will be addressed in future work.

\section{DISCUSSION, LIMITATIONS, and FUTURE WORK}\label{Discussion}

\subsection{Replication of the tag frequency}\label{replicate}
There are several sources of error that should be taken into account while building the MARS tag. First, is the \textit{parasitic reactances}, such as trace, transistor, and inductor capacitance, that can all create a significant effect on the output frequency. Parasitics capacitances near $C_{0}$ of the MCO are emphasized, since small changes in the values of $C_0$ in series with $C_1$ and $C_2$ will result in large changes in $C_{eq}$. Many of these parasitic reactances can be circumvented in the manufacturing stage by usage of tunable traces and passives to achieve the desired frequency. Second, the \textit{quality factor} of the passive components can also determine the accuracy of output frequency. Thus, higher quality passive components should be considered when operating at higher frequencies, e.g., ceramic capacitors made with C0G (NP0) material. Third, humidity and temperature may also play a role in changing passive component values. However, in our indoor testing environment, there have been no observable changes in the passive values due to changes in ambient humidity and temperature between different rooms.

Additionally, there are further difficulties when attempting to calculate the expected frequency of the MCO. First, the value of the capacitor network of $C_{0}$ is hard to determine because it is dependent on $C_{blocking}$, $C_{shift}$, and $C_{JFET}$ which are difficult to measure. Second, the constantly changing gate output of the MCO results in a constantly changing transistor capacitance for the zero-$V_{th}$ MOSFET due to the Miller effect. Changes in $V_{dd}$ and $R_{adjust}$ can result in changes of the offset and range of the gate output oscillation, causing different effective transistor capacitance. Experiments and simulations have shown that they can lead to a small change in the output frequency. As the value for $R_{adjust}$, decreases, the MCO will effectively turn into a ESCO configuration and a sudden jump in oscillation can be observed.

\vspace{-0.1in}
\subsection{Form Factor}\label{sec:formfactor}
With inexpensive, thin, and flexible interface circuits, one can imagine having interfaces in a peelable book, similar to  sticky notes, where an interface can be stuck to a wall, book, or surface wherever it is needed. To work toward this vision, we have tried to adopt a thin form factor when possible, such as a paper-based game controller and post-it note-like wireless microphone. Our photodiode-based prototypes are relatively close to the right form factor, and recent research suggests that our actives, passives \cite{myny2018development, bonnassieux20212021} and thermoelectric generator prototypes could become thinner and more flexible in the near future \cite{ren2021high}. While we have shown some rectangular examples of a swipe-based sensor in \autoref{sec: touch}, the technology enables a large design space (e.g., jogwheels ) for exploring creative touch-based nanopower wireless sensing on objects in the future. 

\vspace{-0.1in}
\subsection{Strategies for increasing operational range}\label{sec:range_discussion}
Different strategies can be employed to increase the operational range. An obvious one is to custom design higher gain antennas that are conformable to the size requirements of a post-it note. Range is a function of frequency at which the transmitter and receiver operate. In the future, the operational range may be considerably increased by operating in a lower frequency FCC free band (e.g., 35MHz rather than 900Mhz). Another possibility for increasing range is to build a Modified Clapp Oscillator tag in the 10s of MHz for active transmission. Shifting from backscatter to active transmission increases range but requires the additional overhead of a bigger power harvester, which may still be practical for certain applications. Backscatter is often dependent on directionality and TX-RX placement. To address this issue we could use a bistatic TX-RX configuration where there is a central omni-directional transmitter and multiple receivers placed strategically around the desired region of operation. This strategy could be useful for larger spaces such as an auditorium. 

\vspace{-0.1in}
\subsection{Placement of Tag}\label{sec:tag_placement}
The operation of each MARS tag and the SNR are affected by the direction of the antenna, as well as when the user accidentally covers the majority of the antenna while interacting with the tag. Interestingly, we did notice that touching the antenna partially can cause an improvement in the SNR. In the future, rather than using a directional antenna, flexible antennas with uniform field may be employed to improve SNR and PESQ scores. Another phenomenon which affects the data quality in case of audio is the user's closeness to the flexible microphone patch. If the user gets too close to the flexible microphone patch, the capacitance of the microphone varies the frequency.

\vspace{-0.1in}
\subsection{Alternate Methods of Power Harvesting}\label{sec:altPH}
Reducing the power to the nanowatt range from the microwatt range  has a direct effect on the type of power harvesters and power management circuits which can be employed in our MARS system.   For example, RF harvesting (e.g., 5G \cite{eid20215g}) might be feasible for perpetually running sensors. In the future, one could also imagine utilizing thinner/more flexible versions of the harvesters, which usually tend to be lower in energy density (e.g., thermoelectric generators \cite{ren2021high}). 

\vspace{-0.1in}
\subsection{Tangible Privacy}\label{sec:privacy}
While the MARS system enables wireless sensing for a variety of input modalities, it also opens research questions about privacy and the user's perception and expectation of privacy. One specific concern is that the technology can be embedded into everyday objects so as to be invisible to the user \cite{ahmad2020tangible}. In the future, we plan to conduct a user study for MARS specifically exploring the privacy aspects.  How can we make changes in the MARS tag design to communicate sensing intention? Are there interaction design patterns, such as requiring the user to place their finger on the thermogenerator button before the microphone will work, that signal to the user when the system is active and what control they have?


\section{CONCLUSION}\label{sec:conclusion}
MARS creates sub $\mu$W sensing tags in flat form factors that can be incorporated on to surfaces such as walls and books or on game controllers.  Frequency shifted backscatter allows many tags to operate in the same region. Using the changing capacitive, inductive, or voltage  properties of  sensors allows direct and highly efficient control of the tag's oscillator for transmission of sensor data. Example applications include battery-free button-based game controllers, sliders for lighting level control, sliders for identification and selection of items from a menu, and remote microphones for auditoriums or home smart speaker systems. With further research, we expect many more interfaces can be developed and, one day, adding controls to a ``smart'' environment will be as simple as placing stickers wherever an interface is needed.

\begin{acks}
We would like to thank Omer Inan for lending us Agilent 34410 without which we would never be able to accomplish this paper. We want to thank Bashima Islam, Tavenner Hall, Purnendu, and Saiganesh Swaminathan for proofreading and giving suggestions on the manuscript. We also thank Canek Fuentes-Hernandez and GT COSMOS lab for brainstorming the application scenarios. Finally, we would like to thank Cisco for funding this research. 
\end{acks}

\bibliographystyle{ACM-Reference-Format}
\bibliography{sample-manuscript}

\appendix
\section{Appendix}
\label{Appendix}

\begin{center}
\vspace{-0.15in}
\begin{figure}[!ht]
\includegraphics[width=\columnwidth]{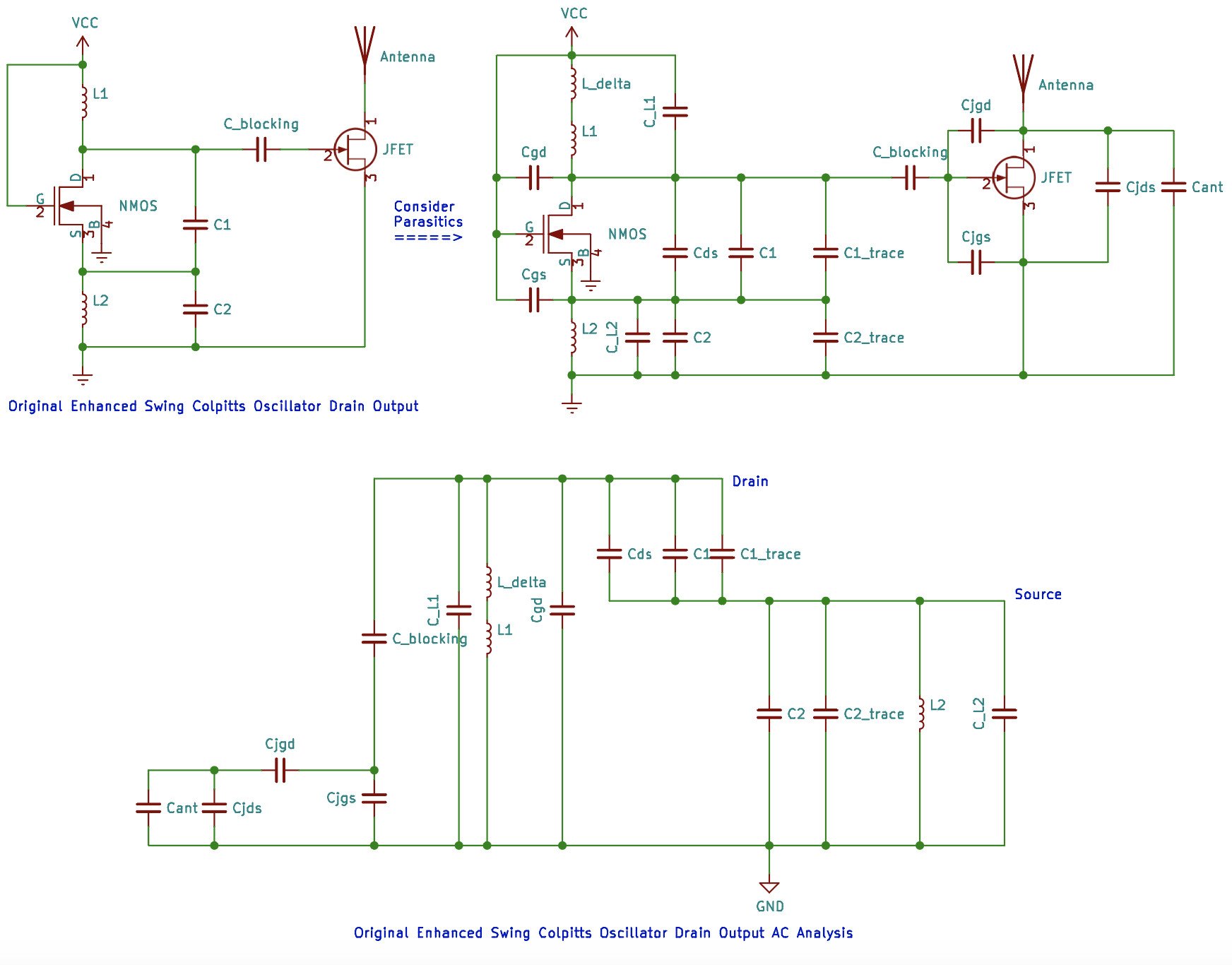}
\caption{ Drain output AC analysis of ESCO oscillator based backscatter} 
\label{fig:drain_AC_analysis}
\end{figure}
\vspace{-0.2in}
\end{center}

\begin{center}
\vspace{-0.15in}
\begin{figure}[!ht]
\includegraphics[width=\columnwidth]{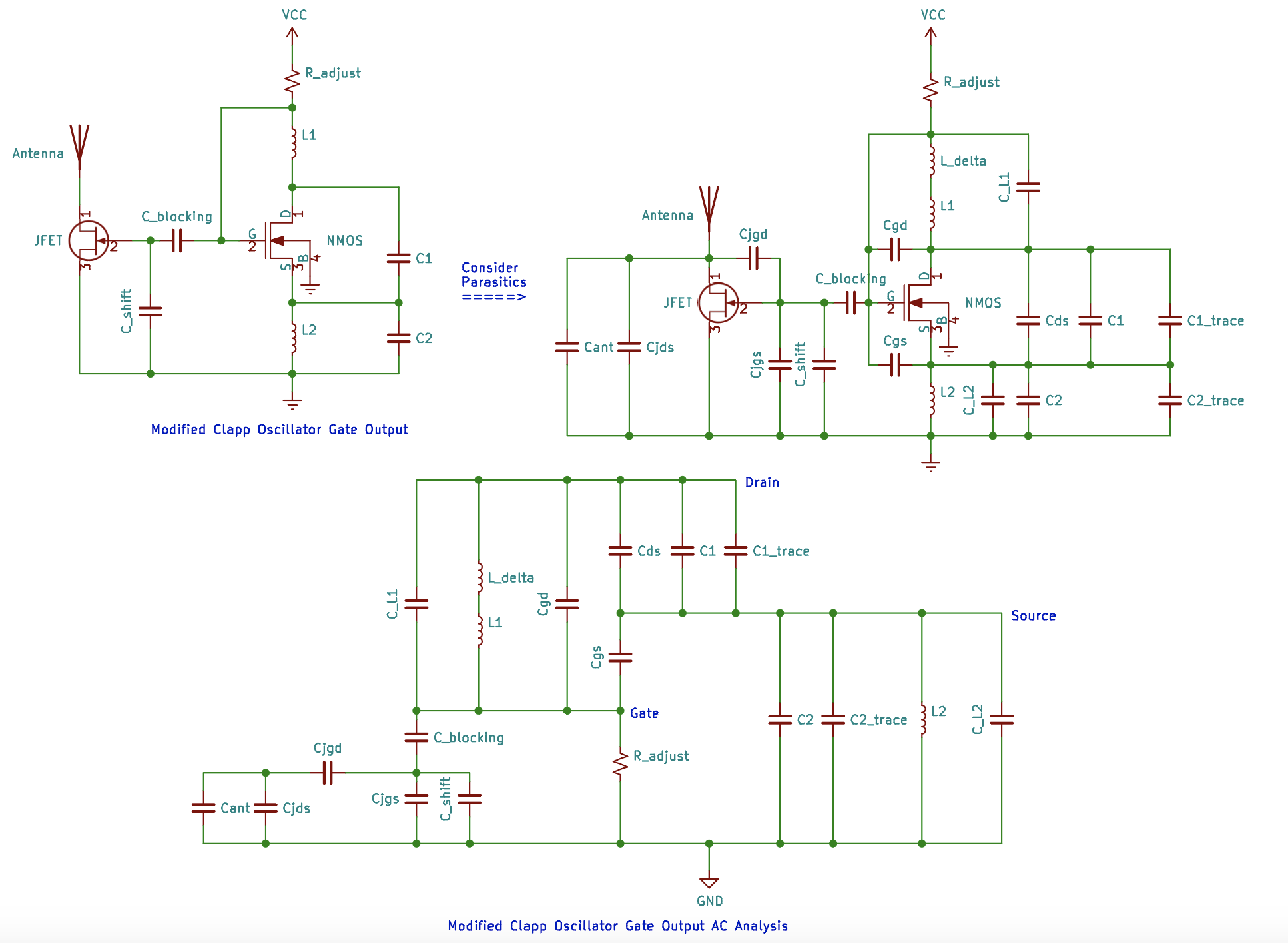}
\caption{ Gate output AC analysis of Modified Clapp Oscillator (MCO) based backscatter} 
\label{fig:gate_AC_analysis}
\end{figure}
\vspace{-0.2in}
\end{center}

\end{document}